\documentclass[useAMS,usenatbib,a4paper]{mn2e}

\usepackage{amssymb}
\usepackage{amsmath}
\usepackage{graphicx}
\usepackage{subfigure}
\usepackage{booktabs}
\usepackage{txfonts}
\usepackage{ifpdf}
\ifpdf
  \usepackage[pdftex,colorlinks=true,breaklinks=true,citecolor=blue]{hyperref}
\else
  \usepackage[ps2pdf,colorlinks=true,citecolor=blue]{hyperref}
\fi
\usepackage{color}

\newif\ifbw
\bwfalse

\newcommand{\eqn}[1]{(#1)}
\newcommand{\Eqn}[1]{(#1)}
\newcommand{\tbl}[1]{Table~#1}

\newcommand{\fig}[1]{Fig.~#1}

\newcommand{\sectn}[1]{Sec.~#1}

\newcommand{\etal}{\mbox{\it et al.}}
\newcommand{\eg}{\mbox{\it e.g.}}
\newcommand{\ie}{\mbox{\it i.e.}}

\newcommand{\kelvin}{{K}}

\newcommand{\frw}{{FRW}}
\newcommand{\frwtext}{{Friedmann-Robertson-Walker}}

\newcommand{\cmb}{{CMB}}
\newcommand{\cmbtext}{{cosmic microwave background}}

\newcommand{\wmap}{{WMAP}}
\newcommand{\wmaptext}{{Wilkinson Microwave Anisotropy Probe}}

\newcommand{\ilctext}{{internal linear combination}}

\newcommand{\bianchi}{{Bianchi}}
\newcommand{\bianchiviih}{{Bianchi VII{$_{\lowercase{h}}$}}}

\newcommand{\lcdm}{\ensuremath{\Lambda}{CDM}}
\newcommand{\lcdmtext}{\ensuremath{\Lambda} Cold Dark Matter}

\newcommand{\healpix}{{\tt HEALPix}}

\newcommand{\anicosmo}{\mbox{\tt ANICOSMO}}

\newcommand{\spcend}{\ensuremath{\:}}

\newcommand{\cconj}{\ensuremath{\ast}}

\newcommand{\sphere}{\ensuremath{{\mathbb{S}^2}}}

\newcommand{\vect}[1]{\ensuremath{\mbox{\boldmath ${#1}$}}}

\newcommand{\dx}{\ensuremath{\mathrm{\,d}}}
\newcommand{\dmu}[1]{\ensuremath{\dx \Omega(#1)}}

\newcommand{\hub}{\ensuremath{H}}
\newcommand{\hubsmall}{\ensuremath{h}}

\newcommand{\Den}{\ensuremath{\Omega}}
\newcommand{\Dentot}{\ensuremath{\Den_{\rm total}}}
\newcommand{\Denmat}{\ensuremath{\Den_{\rm m}}}
\newcommand{\Denmatc}{\ensuremath{\Den_{\rm c}}}
\newcommand{\Denmatb}{\ensuremath{\Den_{\rm b}}}
\newcommand{\Dencurvature}{\ensuremath{\Den_{k}}}

\newcommand{\Denlambda}{\ensuremath{\Den_{\Lambda}}}

\newcommand{\sa}{\ensuremath{\omega}}

\newcommand{\euls}{\ensuremath{\eula, \eulb, \eulc}}
\newcommand{\eula}{\ensuremath{\alpha}}
\newcommand{\eulb}{\ensuremath{\beta}}
\newcommand{\eulc}{\ensuremath{\gamma}}

\newcommand{\el}{\ensuremath{\ell}}
\newcommand{\m}{\ensuremath{m}}

\newcommand{\elmax}{\ensuremath{{L}}}

\newcommand{\p}{\ensuremath{^\prime}}

\newcommand{\kron}[2]{\ensuremath{\delta_{{#1}{#2}}}}

\renewcommand{\exp}[1]{\ensuremath{{\rm e}^{#1}}}
\newcommand{\shfarg}[3]{\ensuremath{Y_{#1#2}({#3})}}
\newcommand{\shfargc}[3]{\ensuremath{Y_{#1#2}^\cconj({#3})}}

\newcommand{\shc}[3]{\ensuremath{{#1}_{{#2}{#3}}}}
\newcommand{\shcc}[3]{\ensuremath{{#1}_{{#2}{#3}}^\cconj}}

\newcommand{\pixarea}{\ensuremath{{\Omega_{\rm{pix}}}}}

\newcommand{\bx}{\ensuremath{x}}
\newcommand{\bhand}{\ensuremath{\kappa}}
\newcommand{\bh}{\ensuremath{h}}

\newcommand{\bsheardim}{\ensuremath{\sigma}}
\newcommand{\bshearonetwo}{\ensuremath{\left(\frac{\sigma_{12}}{H}\right)_0}}
\newcommand{\bshearonethree}{\ensuremath{\left(\frac{\sigma_{13}}{H}\right)_0}}
\newcommand{\bvort}{\ensuremath{\left(\frac{\omega}{H}\right)_0}}
\newcommand{\bshearinline}{\ensuremath{(\sigma/H)_0}}
\newcommand{\bvortinline}{\ensuremath{(\omega/H)_0}}

\newcommand{\prob}{\ensuremath{{\rm P}}}
\newcommand{\given}{\ensuremath{{\,|\,}}}
\newcommand{\evidence}{\ensuremath{{E}}}
\newcommand{\param}{\ensuremath{\Theta}}
\newcommand{\bparam}{\ensuremath{\Theta_{\rm B}}}
\newcommand{\cosmoparam}{\ensuremath{\Theta_{\rm C}}}
\newcommand{\anisoparam}{\ensuremath{\Theta_{\rm A}}}
\newcommand{\fitdata}{\ensuremath{\vect{d}}}

\newcommand{\fitmodelsel}{\ensuremath{M}}

\newcommand{\fittmpl}{\ensuremath{\vect{b}}}

\newcommand{\fitdatasky}{\ensuremath{d}}
\newcommand{\fitdataalm}{\ensuremath{\shc{d}{\el}{\m}}}
\newcommand{\fittmplalm}{\ensuremath{\shc{b}{\el}{\m}}}
\newcommand{\fitdataalz}{\ensuremath{\shc{d}{\el}{0}}}
\newcommand{\fittmplalz}{\ensuremath{\shc{b}{\el}{0}}}
\newcommand{\mnoise}{\ensuremath{m}}

\renewcommand{\exp}[1]{\ensuremath{{\rm exp}{#1}}}
\renewcommand{\elmax}{\ensuremath{{\el_{\rm max}}}}
\renewcommand{\eqn}[1]{Eqn.~(#1)}
\renewcommand{\Eqn}[1]{Eqn.~(#1)}

\title[\bianchiviih\ and \wmap]
   {Bayesian analysis of anisotropic cosmologies: \bianchiviih\ \\and \wmap}

\author[McEwen \etal]
  {J.~D.~McEwen$^{1,2}$\thanks{jason.mcewen@ucl.ac.uk},
   T.~Josset$^{1,3,4}$\thanks{tjosset@ens-cachan.fr}, 
   S.~M.~Feeney$^1$,\thanks{stephen.feeney.09@ucl.ac.uk}, 
   H.~V.~Peiris$^1$\thanks{h.peiris@ucl.ac.uk}  
   and A.~N.~Lasenby$^5$\thanks{a.n.lasenby@mrao.cam.ac.uk} \\ 
  ${}^1$Department of Physics and Astronomy, 
    University College London, London WC1E 6BT, U.K.\\
  ${}^2$Mullard Space Science Laboratory (MSSL), 
    University College London, Surrey RH5 6NT, U.K.\\
  ${}^3$Ecole Normale Sup\'{e}rieure de Cachan, 94230 Cachan, France\\
  ${}^4$Universit\'{e} Pierre et Marie Curie, 75005 Paris, France\\
  ${}^5$Astrophysics Group, Cavendish Laboratory, Madingley Road,
    Cambridge CB3 0HE, U.K.
}

\date{Accepted ---. Received ---; in original form ---}
\pagerange{\pageref{sec:intro}--\pageref{lastpage}} 
\pubyear{2013}

\def\LaTeX{L\kern-.36em\raise.3ex\hbox{a}\kern-.15em
    T\kern-.1667em\lower.7ex\hbox{E}\kern-.125emX}

\begin{document}
\maketitle

\begin{abstract}
  We perform a definitive analysis of \bianchiviih\ cosmologies with
  \wmap\ observations of the \cmbtext\ (\cmb) temperature
  anisotropies.  Bayesian analysis techniques are developed to study
  anisotropic cosmologies using full-sky and partial-sky, masked \cmb\
  temperature data.  We apply these techniques to analyse the full-sky
  \ilctext\ (ILC) map and a partial-sky, masked W-band map of \wmap\
  9-year observations.  In addition to the physically motivated
  \bianchiviih\ model, we examine phenomenological models considered
  in previous studies, in which the \bianchiviih\ parameters are
  decoupled from the standard cosmological parameters. In the two
  phenomenological models considered, Bayes factors of 1.7 and 1.1
  units of log-evidence favouring a Bianchi component are found in
  full-sky ILC data.  The corresponding best-fit Bianchi maps
  recovered are similar for both phenomenological models and are very
  close to those found in previous studies using earlier \wmap\ data
  releases.  However, no evidence for a phenomenological Bianchi
  component is found in the partial-sky W-band data.  In the physical
  \bianchiviih\ model we find no evidence for a Bianchi component:
  \wmap\ data thus do not favour \bianchiviih\ cosmologies over the
  standard \lcdmtext\ (\lcdm) cosmology.  It is not possible to
  discount \bianchiviih\ cosmologies in favour of \lcdm\ completely,
  but we are able to constrain the vorticity of physical \bianchiviih\
  cosmologies at $\bvortinline < 8.6 \times 10^{-10}$ with 95\%
  confidence.
\end{abstract}


\section{Introduction}
\label{sec:intro}


The \lcdmtext\ (\lcdm) cosmological concordance model has recently
emerged as an accurate description of our Universe.  In this model
the current Universe is dominated by dark energy and dark matter, with
structure seeded by primordial density perturbations generated 
during an inflationary phase in the early Universe. Support for \lcdm\
is derived from a range of recent high-precision cosmological observations, 
with measurements of the \cmbtext\ (\cmb), in particular those made by the 
\wmaptext\ (\wmap) \citep{komatsu:2010,larson:2011,hinshaw:2013}, 
playing a leading role.
While the concordance model is undeniably successful, it is important to 
test the assumptions on which it is based.  One of the most fundamental
assumptions of \lcdm\ cosmology is the cosmological principle; namely,
that the Universe is homogenous and isotropic on large scales. Evidence 
that this assumption is inaccurate would necessitate revision of \lcdm.

In this article we seek to test the global isotropy of the Universe.
When studying phenomena beyond cosmological concordance it is
important to do so in the context of a well-motivated cosmological
model.  We focus on the homogenous but anisotropic Bianchi models.  In
these models the assumption of isotropy about each point in the
Universe is relaxed, yielding more general solutions to Einstein's
field equations.  For small anisotropy, as demanded by current
observations, linear perturbation about the standard \frwtext\ (\frw)
metric may be applied, leading to a subdominant,
deterministic contribution to the \cmb\ fluctuations.  In this setting
\cmb\ fluctuations may be viewed as the sum of a deterministic Bianchi
contribution and the usual stochastic contribution that arises in the
\lcdm\ model.

\newpage

The induced \cmb\ temperature fluctuations that result in the
homogenous Bianchi models were first studied by \cite{collins:1973}
and \cite{barrow:1985} (and subsequently \citealt{barrow:1986}),
ignoring the effects of dark energy.  \cite{barrow:1985} focus on the
most general Bianchi VII${}_\bh$ and IX types, corresponding to
open/flat and closed universes respectively.  Since tight constraints
have already been placed on Bianchi IX models by \cite{barrow:1985},
we follow the recent trend in the literature and focus on the
open/flat \bianchiviih\ models, in which geodesic focusing produces
spiral-type contributions in the \cmb.  Although we focus on \cmb\
temperature fluctuations only in this article, we note that the \cmb\
polarisation contributions induced in Bianchi models have also been
studied recently \citep{pontzen:2007,pontzen:2009,pontzen:2011}.

\bianchiviih\ models were first compared to COBE data by
\cite{bunn:1996} and \cite{kogut:1997}, and to \wmap\ data by
\cite{jaffe:2005,jaffe:2006a}.  A statistically significant
correlation between one of the \bianchiviih\ models and the \wmap\
internal linear combination (ILC) map (\citealt{bennett:2003a}) was
discovered by \cite{jaffe:2005} by modelling the CMB as an unknown
Bianchi component on top of the (fixed) best-fit \wmap\ \lcdm\
cosmology. However, it was noted that the parameters of the best-fit
Bianchi component were incompatible with those of \lcdm.
Nevertheless, quite remarkably it was found that when the \wmap\ data
were `corrected' for the best-fit Bianchi map, some of the so-called
`anomalies' reported in \wmap\ data disappeared
\citep{jaffe:2005,jaffe:2006a,cayon:2006,mcewen:2006:bianchi}. A
modified template-fitting technique was performed by \cite{lm:2006}
and, although a statistically significant template fit was not
reported, the corresponding `corrected' \wmap\ data were again free of
many large scale `anomalies'.  Subsequently,
  \citet{ghosh:2007} used the bipolar power spectrum of \wmap\ data to
  constrain the amplitude of any Bianchi component in the \cmb.

Following this renewed interest in Bianchi models, the \cmb\
temperature fluctuations induced in \bianchiviih\ models incorporating
dark energy were derived by \citet{jaffe:2006b} and \citet{bridges:2006b}
(and subsequently by \citet{pontzen:2007}, \citet{pontzen:2009}
  and \citet{pontzen:2011}, where recombination is treated in a more
  sophisticated manner and reionisation is supported). In this
scenario, a degeneracy between the matter and dark energy densities,
\Denmat\ and \Denlambda\ respectively, is introduced, but the
cosmological parameters of the best-fit Bianchi template found in
\wmap\ data nevertheless remain inconsistent with constraints from the
\cmb\ alone \citep{jaffe:2006c,jaffe:2006b}. Furthermore,
  \cite{pontzen:2007} compared the polarisation power spectra of the
  best-fit \bianchiviih\ model found by \cite{jaffe:2006c} with the
  \textit{WMAP} 3-year data \citep{page:2007} and also concluded that
  the model could be ruled out since it produced greater
  polarization than observed in the \textit{WMAP} data.  A Bayesian
analysis of \bianchiviih\ models was performed by \cite{bridges:2006b}
using \wmap\ ILC data to explore the joint cosmological and Bianchi
parameter space via Markov chain Monte Carlo sampling.  The
\mbox{$\Denmat$--$\Denlambda$} degeneracy of \bianchiviih\ models was
studied more thoroughly and, although a similar best-fit Bianchi
template was found in \wmap\ data, it was again determined that the
parameters of the resulting Bianchi cosmology were inconsistent with
standard constraints.  In a following study by
\cite{bridges:2007:bianchi} it was suggested that the \cmb\ `cold
spot' \citep{vielva:2004,cruz:2006a,vielva:2010} could be driving the
best-fit Bianchi component found in \wmap\ data.

In this article we perform a definitive study of \bianchiviih\
cosmologies with \wmap\ temperature data. We develop a Bayesian
analysis technique capable of studying Bianchi models using
partial-sky \cmb\ observations, allowing us for the first time to
study individual \wmap\ bands rather than the ILC map, which was not
originally intended for cosmological analysis
\citep{bennett:2003a}. We sample the complete set of parameters
describing the \lcdm\ cosmology and the \bianchiviih\ model
simultaneously, and perform the first rigorous study of the physically
motivated scenario where the parameters of the Bianchi model are
coupled to the standard cosmology. To make comparisons with previous
work, we also consider the non-physical scenario where the Bianchi
parameters are decoupled from the standard cosmological parameters. We
employ the latest (and final) 9-year release of \wmap\ observations
(previous studies have considered \wmap\ 1- and 3-year data only), and
perform a Bayesian model-selection analysis of \bianchiviih\ models,
using nested sampling methods
\citep{skilling:2004,feroz:multinest1,feroz:multinest2}, to determine
whether \wmap\ data suggest we inhabit an anisotropic \bianchiviih\
universe instead of a standard isotropic \lcdm\ universe.

The remainder of this article is structured as follows.  In
\sectn{\ref{sec:bianchi_cosmologies}} we briefly review \bianchiviih\
cosmologies.  We describe the Bayesian analysis techniques
that we develop to study full- and partial-sky \cmb\ observations
in \sectn{\ref{sec:bayesian_analysis}}.  Although focus is given to
Bianchi models, these techniques are generic and may be used to study
other general anisotropic cosmologies. In
\sectn{\ref{sec:wmap_analysis}} we apply our analysis to \wmap\ 9-year
data to constrain \bianchiviih\ cosmologies.  Concluding remarks are
made in \sectn{\ref{sec:conclusions}}.

\section{Bianchi VII${}_{\lowercase{h}}$ cosmologies}
\label{sec:bianchi_cosmologies}

\newlength{\plotwidth}
\setlength{\plotwidth}{0.32\columnwidth}

\begin{figure}
\centering

\mbox{
  \includegraphics[bb= 0 40 800 440,clip=,width=\plotwidth]{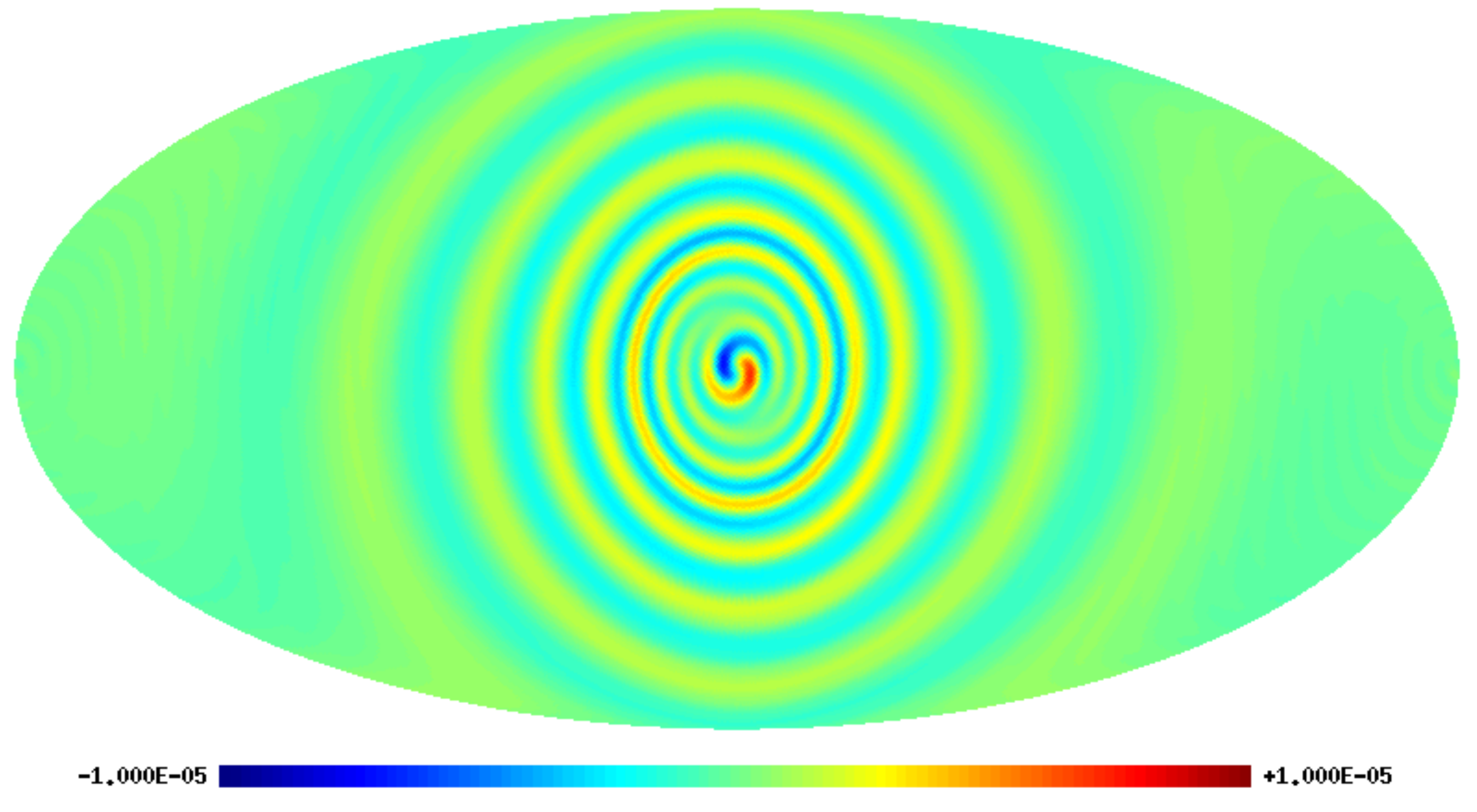}
  \includegraphics[bb= 0 40 800 440,clip=,width=\plotwidth]{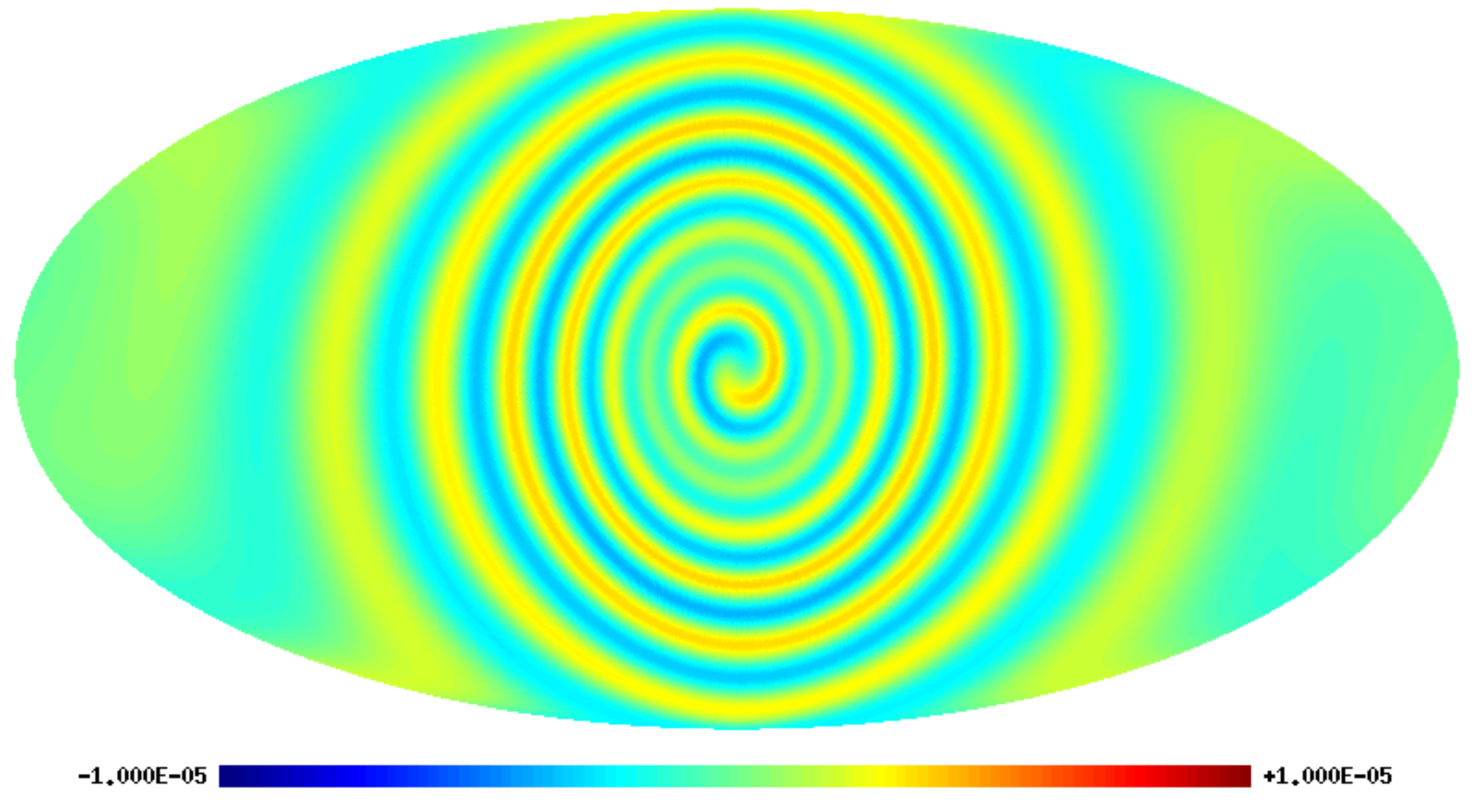} 
  \includegraphics[bb= 0 40 800 440,clip=,width=\plotwidth]{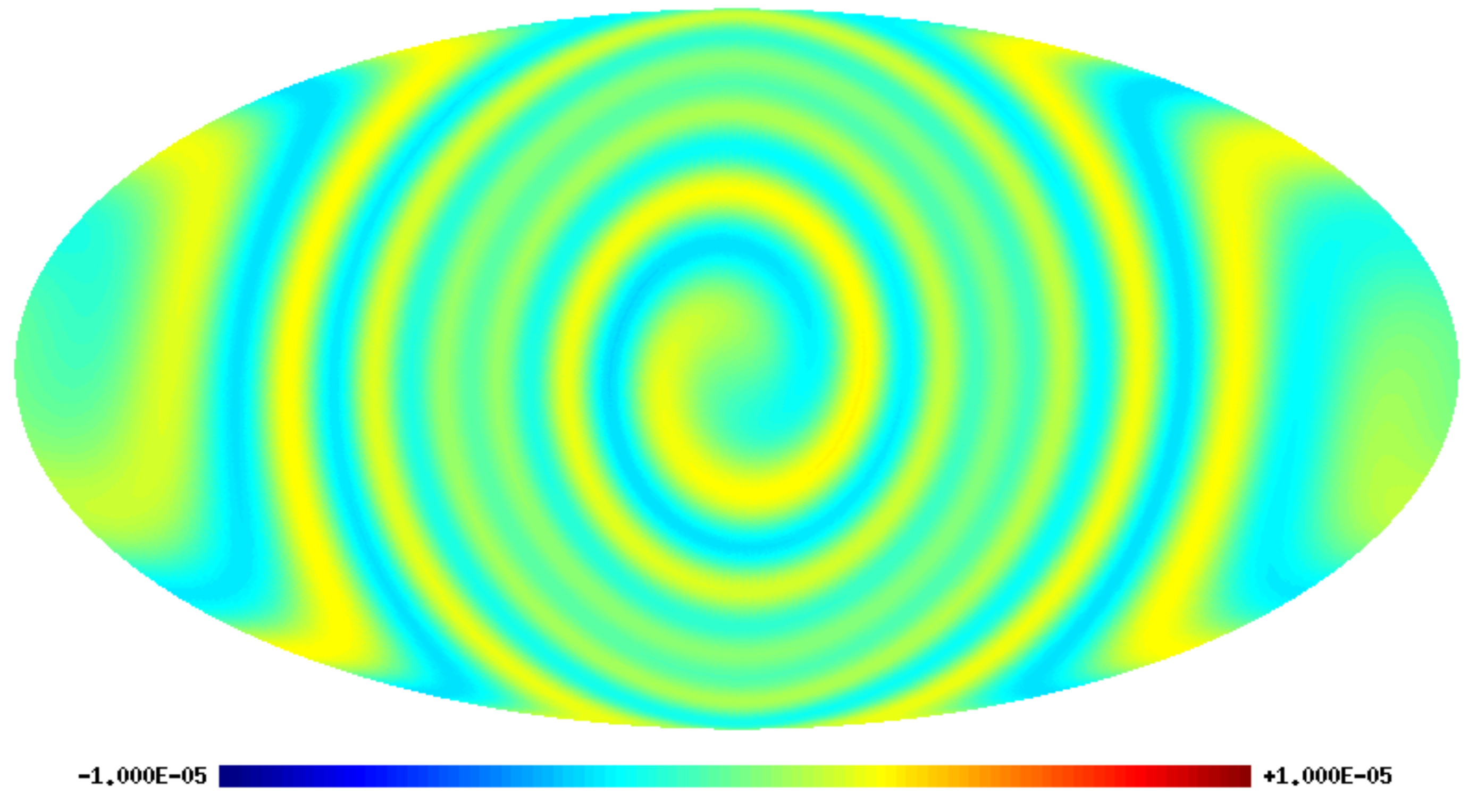}
}
\mbox{
  \includegraphics[bb= 0 40 800 440,clip=,width=\plotwidth]{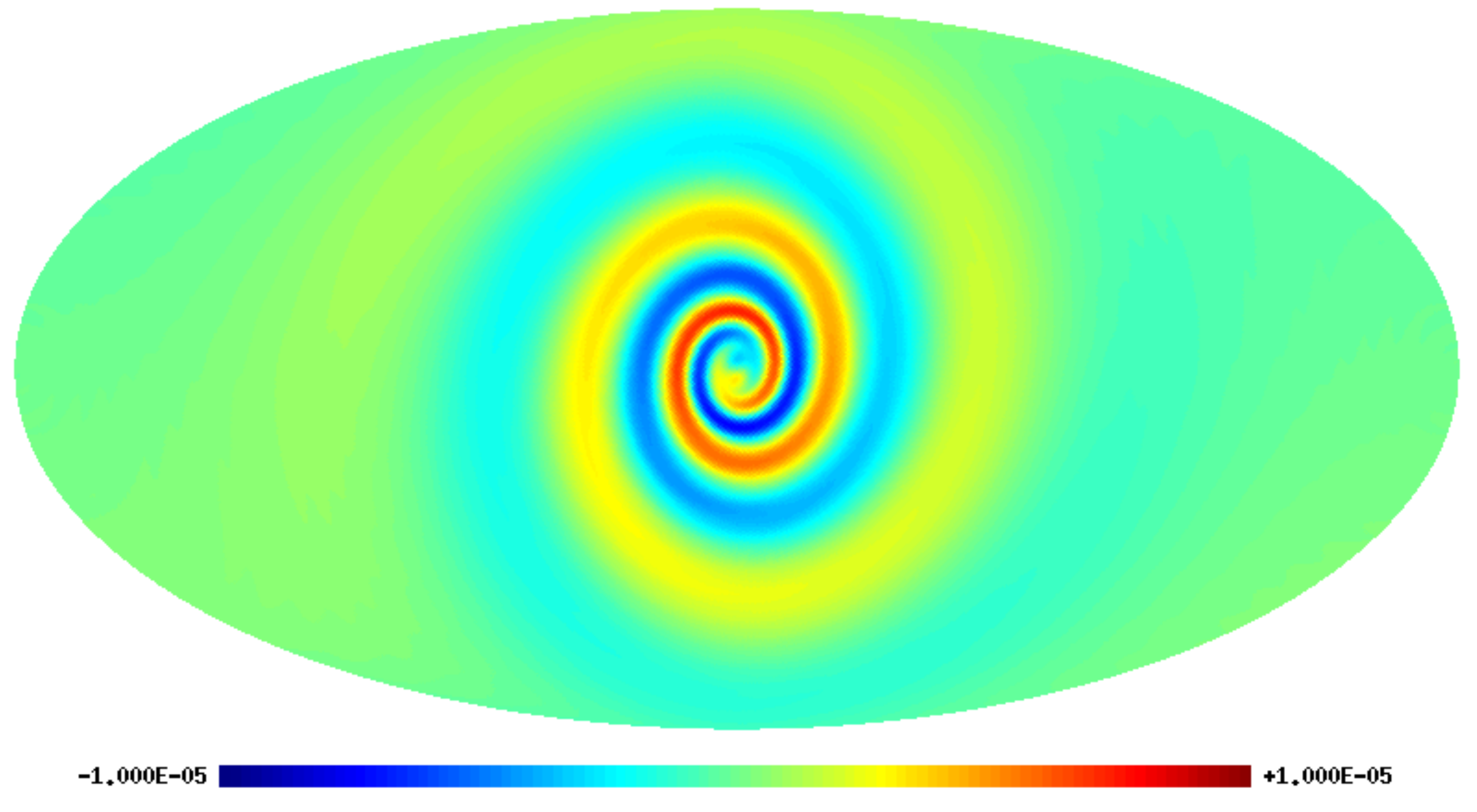} 
  \includegraphics[bb= 0 40 800 440,clip=,width=\plotwidth]{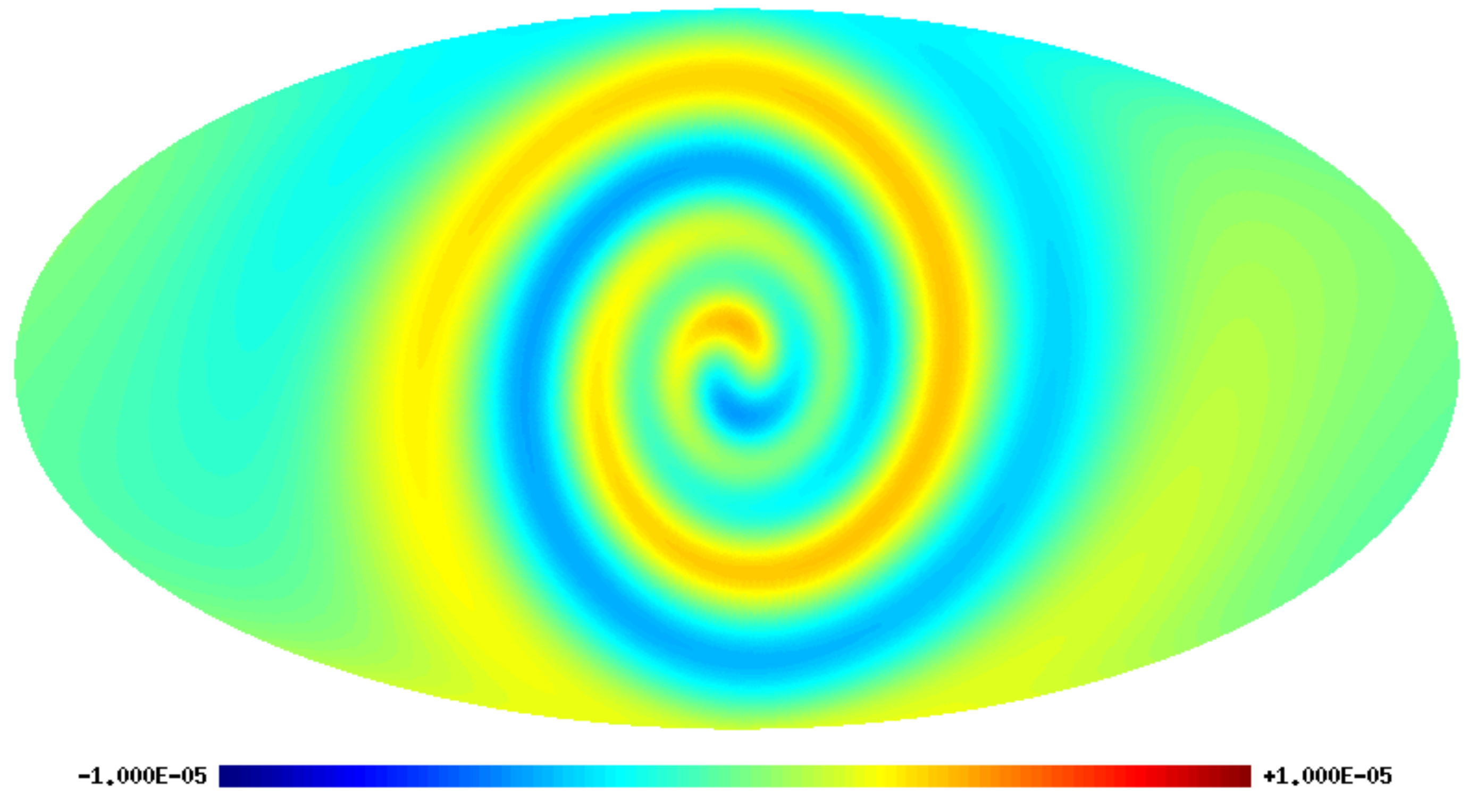} 
  \includegraphics[bb= 0 40 800 440,clip=,width=\plotwidth]{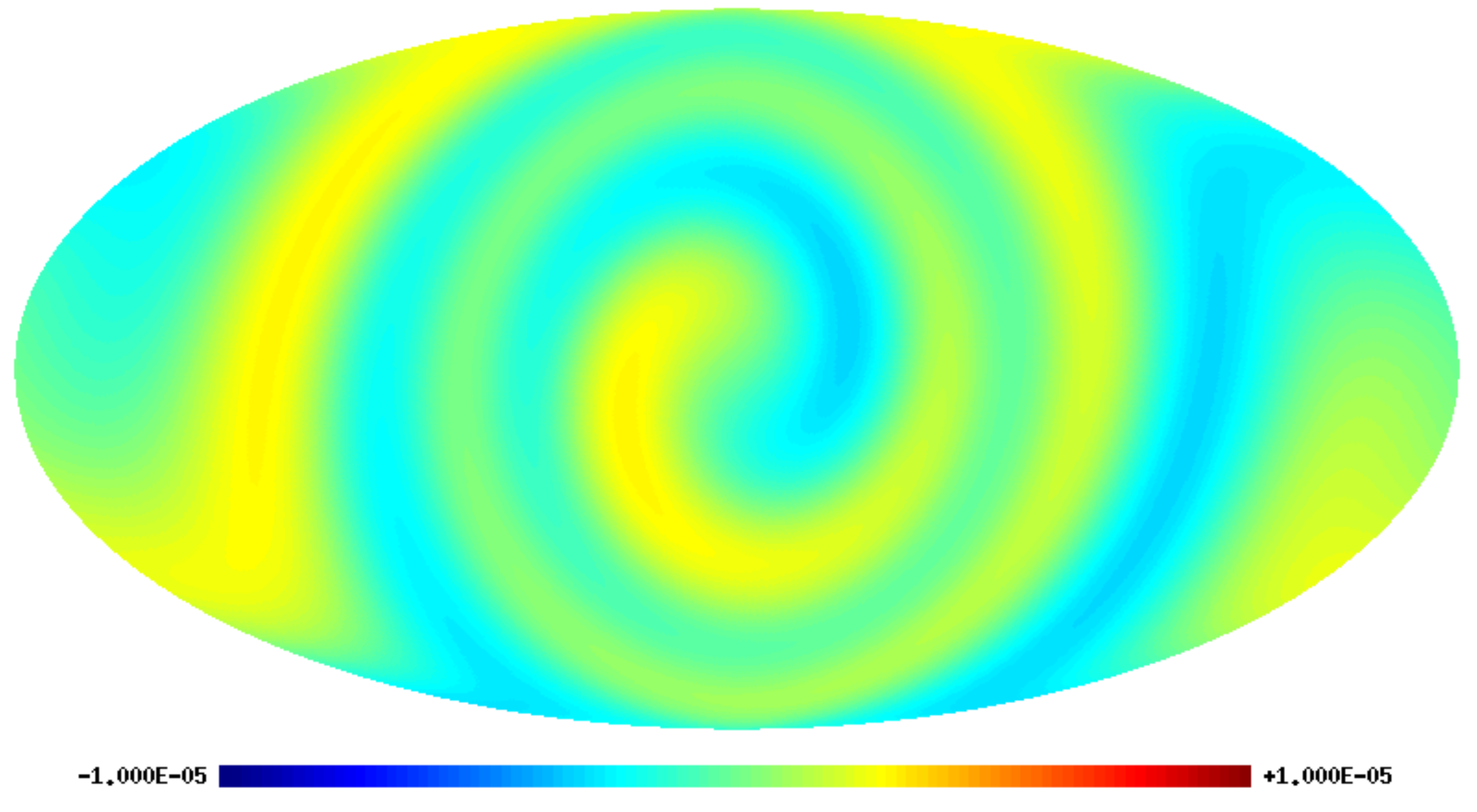}
}
\mbox{
  \includegraphics[bb= 0 40 800 440,clip=,width=\plotwidth]{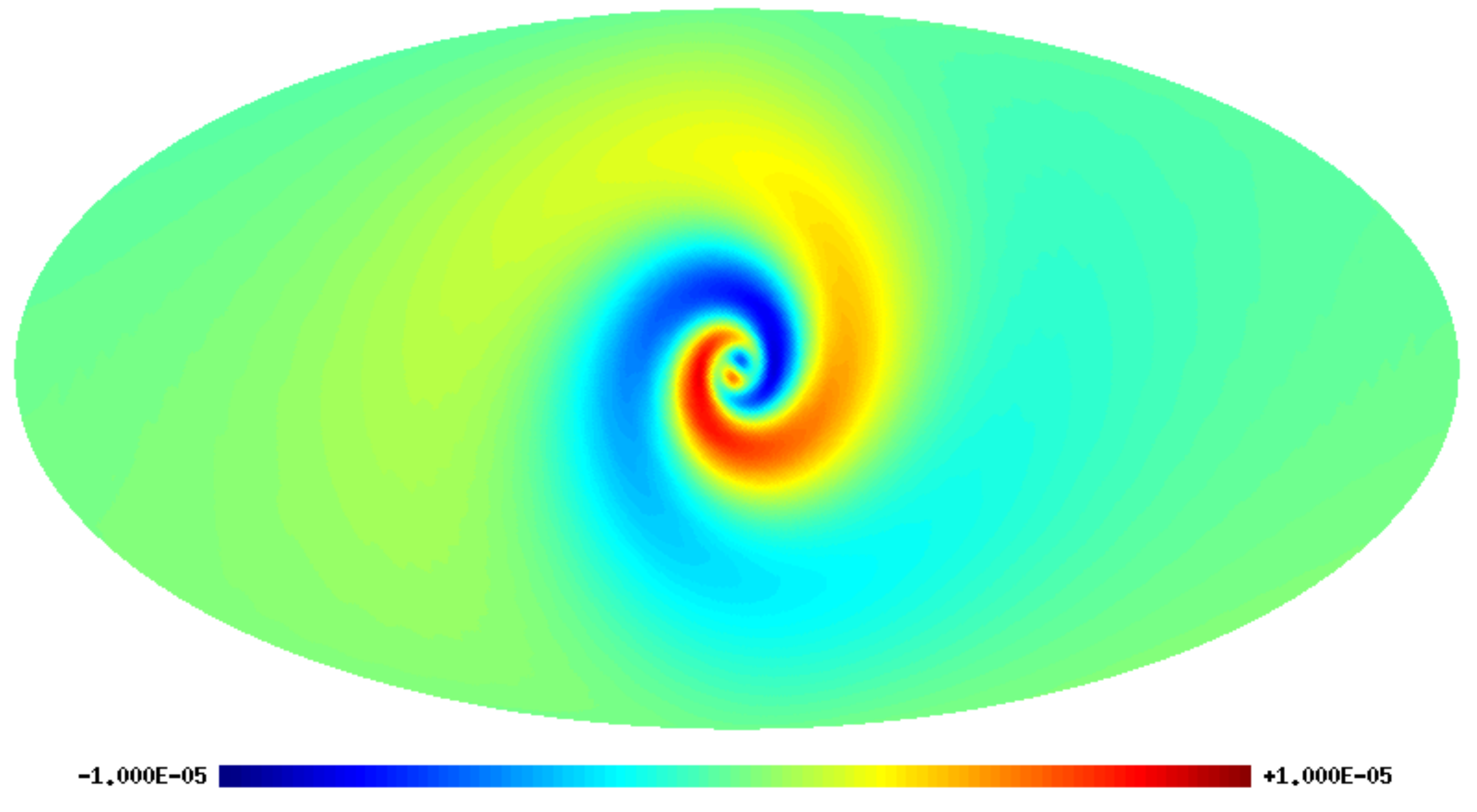} 
  \includegraphics[bb= 0 40 800 440,clip=,width=\plotwidth]{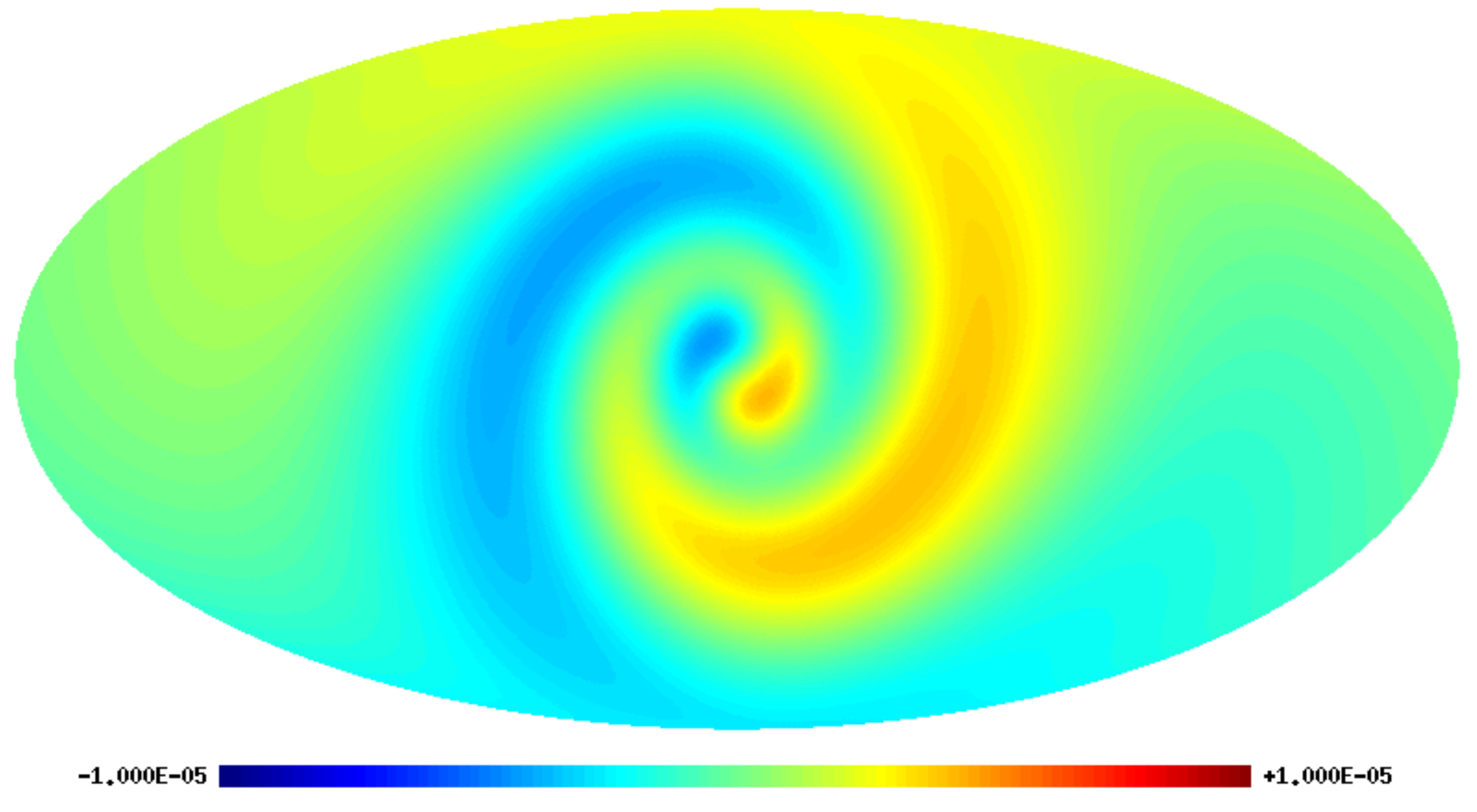} 
  \includegraphics[bb= 0 40 800 440,clip=,width=\plotwidth]{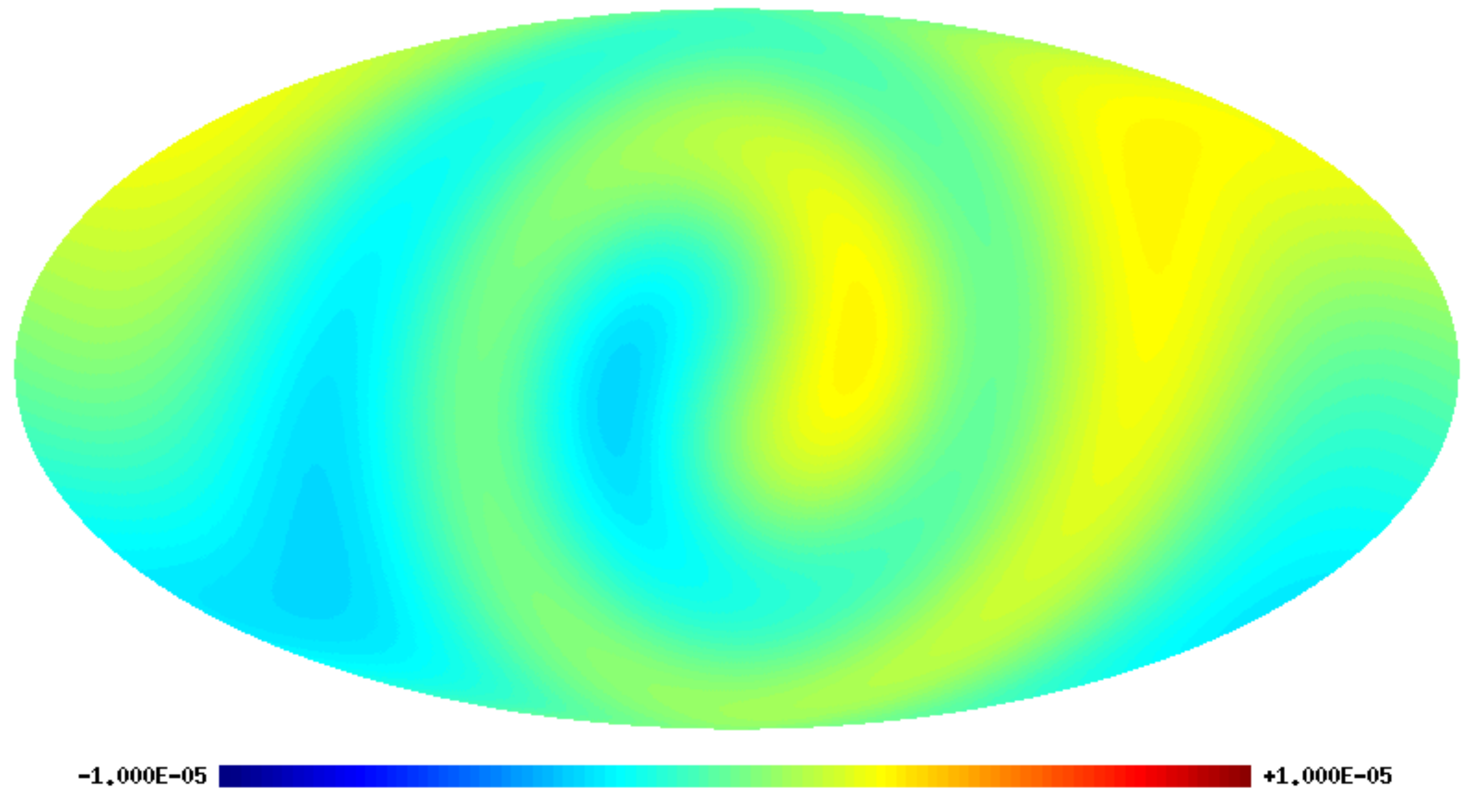}
}
\mbox{
  \includegraphics[bb= 0 40 800 440,clip=,width=\plotwidth]{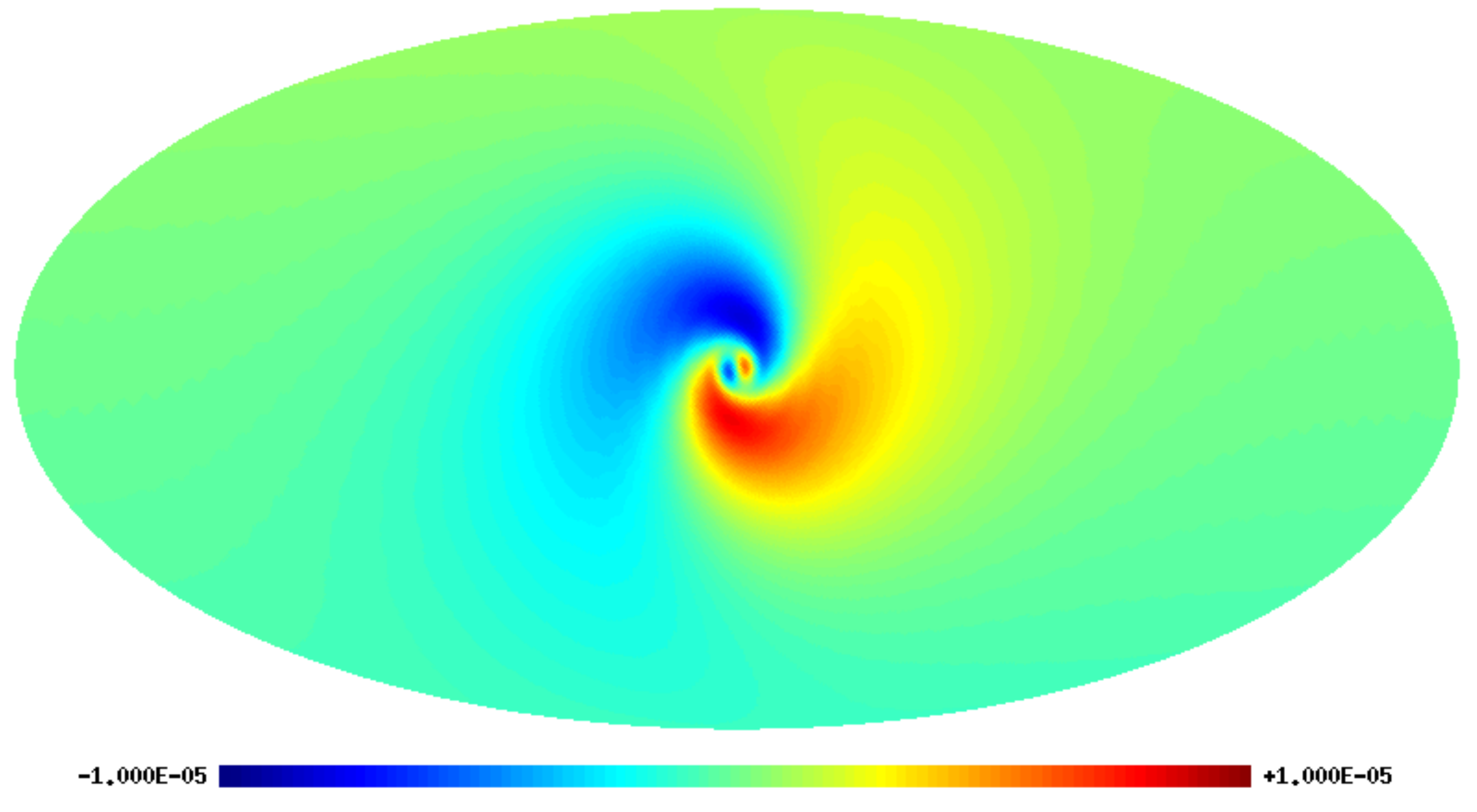} 
  \includegraphics[bb= 0 40 800 440,clip=,width=\plotwidth]{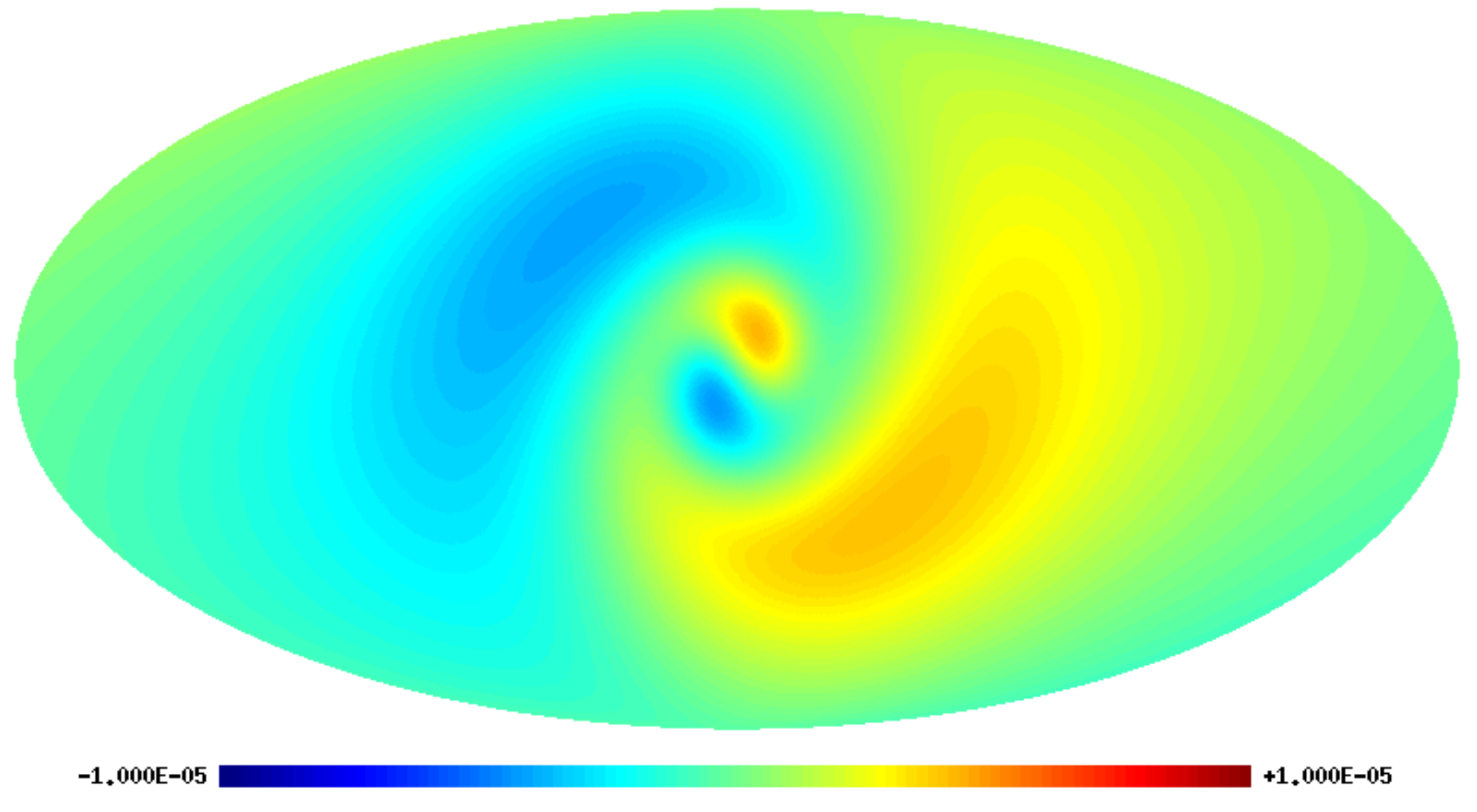} 
  \includegraphics[bb= 0 40 800 440,clip=,width=\plotwidth]{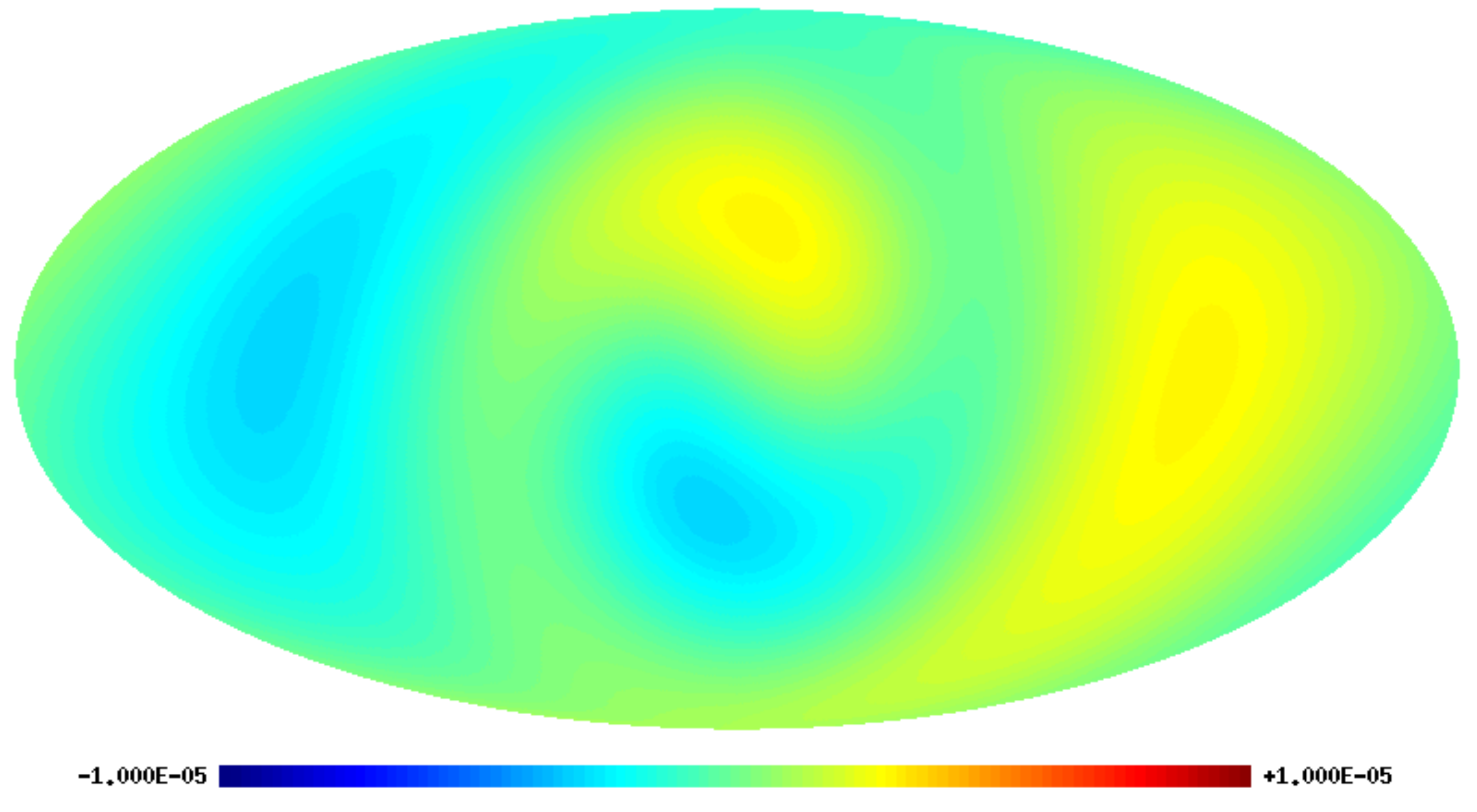}
}
\mbox{
  \includegraphics[bb= 0 40 800 440,clip=,width=\plotwidth]{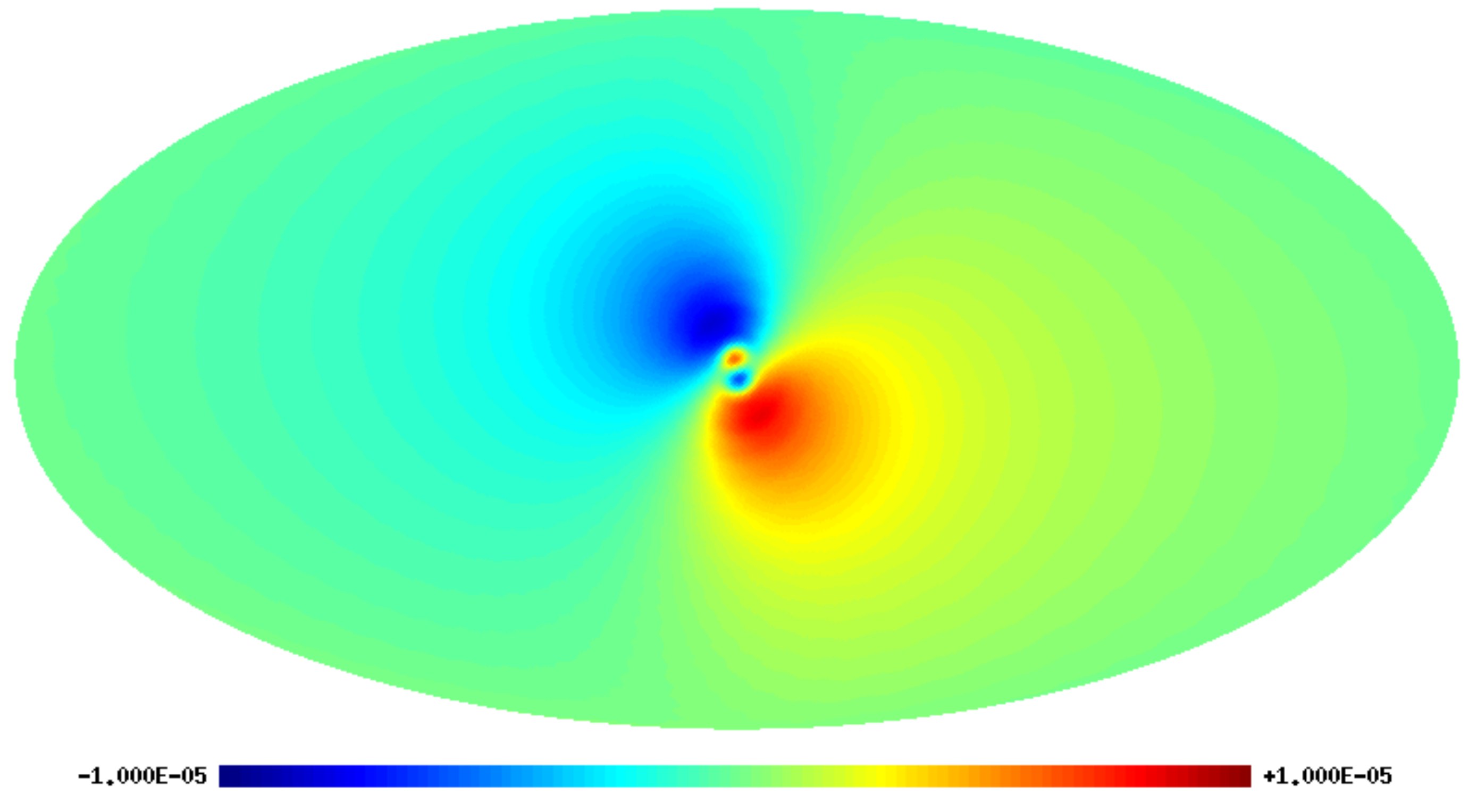} 
  \includegraphics[bb= 0 40 800 440,clip=,width=\plotwidth]{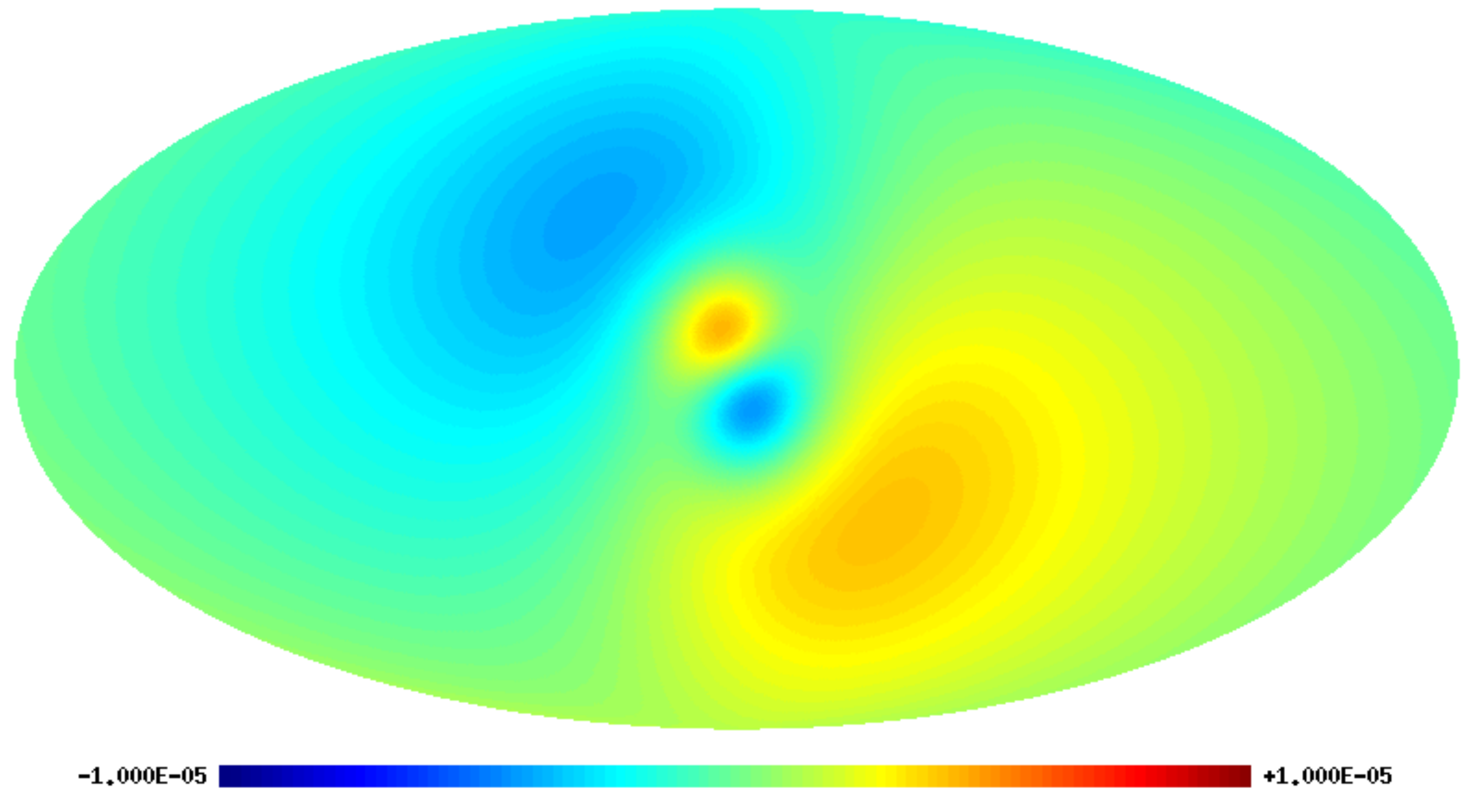} 
  \includegraphics[bb= 0 40 800 440,clip=,width=\plotwidth]{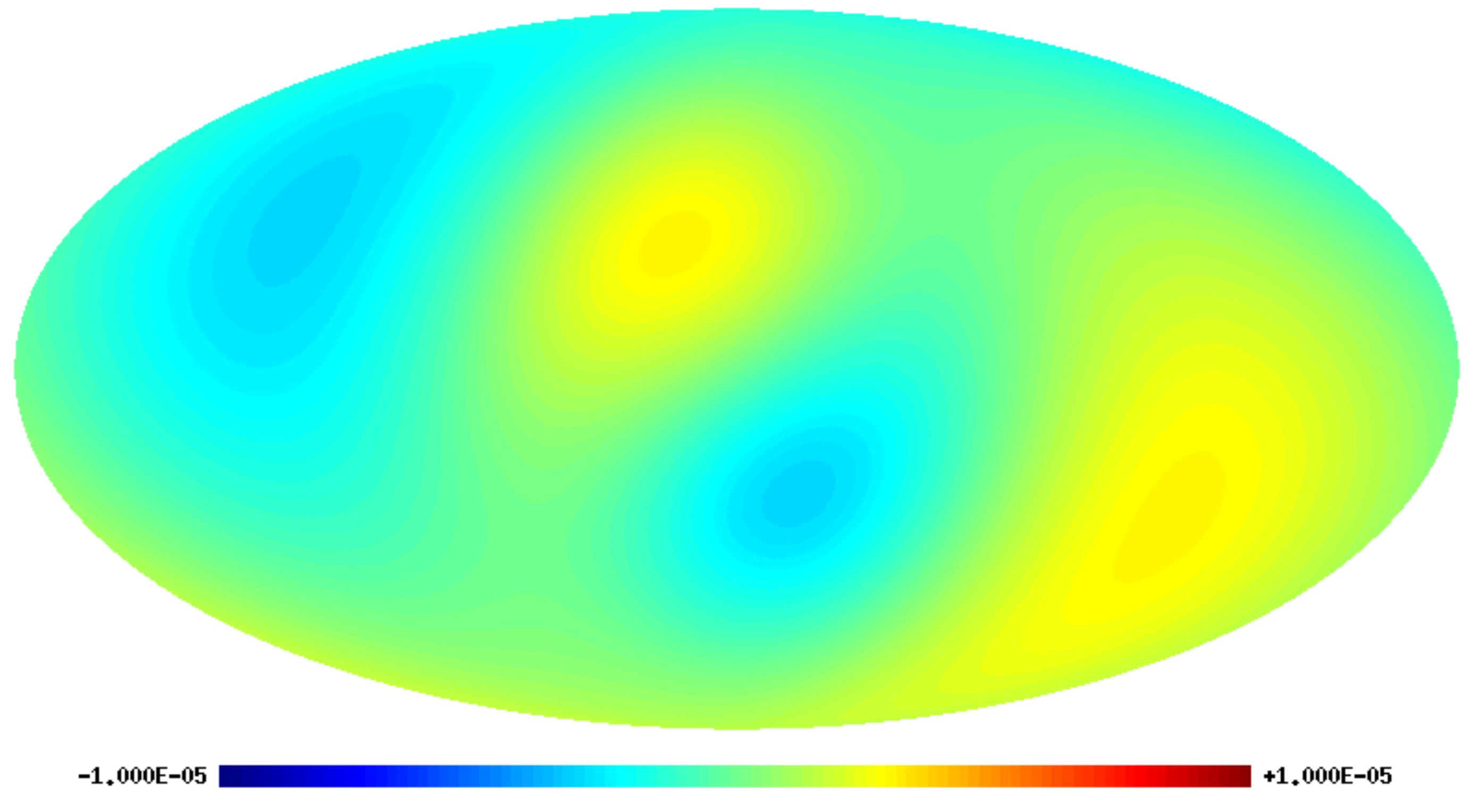}
}
\caption[Simulated \bianchiviih\ maps]{Simulated deterministic \cmb\
  temperature contributions in \bianchiviih\ cosmologies for varying
  $\bx$ and $\Dentot$ (left-to-right \mbox{$\Dentot \in \{0.10, 0.50,
  0.95\}$}; top-to-bottom $\bx\in\{0.1, 0.3, 0.7, 1.5, 6.0 \}$).  In
  these maps the spiral pattern typical of \bianchiviih\ induced
  temperature fluctuations is rotated from the South pole to the
  Galactic centre for illustrational purposes. The parameter values
  $\bhand=+1$ and $(\euls)=(0^\circ,-90^\circ,0^\circ)$ were also set
  when generating these simulations.  }
\label{fig:bianchi_sims}
\end{figure}

In this section we review \bianchiviih\ cosmologies, focusing on their
description; we defer technical details to \cite{barrow:1985} and
\cite{jaffe:2006b}.  Note that we adopt the solutions to the
temperature fluctuations induced in \bianchiviih\ models when
incorporating dark energy, as derived by Lasenby and
also adopted by \cite{bridges:2006b}.

\bianchiviih\ models contain a free parameter, usually denoted \bx, as
first identified by \cite{collins:1973}.  This parameter is related to
the \bh\ parameter of type VII${}_\bh$ models by
\begin{equation}
\bx = \sqrt{\frac{\bh}{1-\Dentot}}
\spcend ,
\end{equation}
where the total energy density $\Dentot = \Denmat + \Denlambda$.
Physically, \bx\ is related to the characteristic wavelength over
which the principle axes of shear and rotation change orientation.
Consequently, \bx\ acts to change the `tightness' of the spiral-type
\cmb\ temperature contributions that are typical of \bianchiviih\
cosmologies.  \bianchiviih\ models are also described by their shear
modes $(\bsheardim_{12}/\hub)_0$ and
  $(\bsheardim_{13}/\hub)_0$, vorticity $\bvortinline$, and
handedness \bhand, where \hub\ is the Hubble parameter.  The
handedness parameter takes the values $\bhand=+1$ and $\bhand=-1$ for
maps with right- and left-handed spirals respectively.  Vorticity is
related to the other parameters by \citep{barrow:1985}
\begin{equation}
\label{eqn:vort}
\bvort= \frac{(1+\bh)^{1/2} (1+9\bh)^{1/2}}{6\,\bx^2\Dentot} 
\sqrt{\bshearonetwo^2 + \bshearonethree^2}
\spcend .
\end{equation}

The spherical harmonic coefficients of the \bianchiviih\ induced
  temperature component are proportional to \mbox{$[(\bsheardim_{12} \pm i
    \bsheardim_{13})/H]_0 $} and are non-zero for azimuthal modes $m=\mp
  1$ only \citep{barrow:1985, mcewen:2006:bianchi,
    pontzen:2007}. Hence, varying the phase of
  $\bsheardim_{12}+i\bsheardim_{13}$ corresponds to an azimuthal rotation,
  \ie\ a change of coordinates, while the rotationally invariant part
  depends on $\bsheardim_{12}^2+\bsheardim_{13}^2$, and we are thus free to
  choose equality of shear modes $\bsheardim=\bsheardim_{12}=\bsheardim_{13}$
  \citep{pontzen:2007}, which we do for consistency with previous
  studies (\eg\ \citealt{jaffe:2005}).
The amplitude of deterministic \bianchiviih\ temperature maps may be
characterised by either $\bshearinline$ or $\bvortinline$ since these
parameters influence the amplitude of the map only and not its
morphology.  

Since the \cmb\ temperature fluctuations induced in \bianchiviih\
models are anisotropic on the sky, the orientation of the \cmb\
contribution may vary also, introducing three additional
degrees-of-freedom.  The orientation of a map representing the
\bianchiviih\ \cmb\ fluctuations is described by the Euler
angles\footnote{The active $zyz$ Euler convention is adopted,
  corresponding to the rotation of a physical body in a {\it{fixed}}
  coordinate system about the $z$, $y$ and $z$ axes by $\gamma$,
  $\beta$ and $\alpha$ respectively.}  $(\euls)$, where for
$(\euls)=(0^\circ,0^\circ,0^\circ)$ the spiral pattern typical of
\bianchiviih\ temperature fluctuations is centred on the South pole.
To summarise, \bianchiviih\ models may be described by the parameter
vector $\bparam = (\Denmat,\, \Denlambda,\, \bx,\, \bvortinline,\, 
\euls)$ (note that we do not include the handedness parameter $\bhand$
in \bparam\ since this is used to distinguish between left- and
right-handed models).

In the analysis performed herein the {\tt
  BIANCHI2}\footnote{\url{http://www.jasonmcewen.org/}} code is used
to simulate the temperature fluctuations induced in \bianchiviih\
models.  \bianchiviih\ models induce only large scale temperature
fluctuations in the \cmb\ and, consequently, the resulting Bianchi
maps have a low band-limit both globally and azimuthally, \ie\ in both
\el\ and \m\ in spherical harmonic space; indeed, only those harmonic
coefficients with $\m=\pm1$ are non-zero \citep{barrow:1985}.  In the
{\tt BIANCHI2} code \bianchiviih\ temperature fluctuations may be
computed directly in either real or harmonic space.  In the analysis
performed herein we compute temperature fluctuations directly in
harmonic space to avoid any pixelisation artefacts and since the
resulting temperature fluctuations can be rotated accurately and
efficiently in harmonic space (see \citealt{mcewen:2006:bianchi}) due
to their low azimuthal band-limit.  Examples of simulated
\bianchiviih\ temperature maps are illustrated in
\fig{\ref{fig:bianchi_sims}} for a range of parameters.

\section{Bayesian analysis of anisotropic cosmologies}
\label{sec:bayesian_analysis}

We describe in this section a generic methodology to perform a
Bayesian analysis of anisotropic cosmologies using \cmb\ observations.
Since we are motivated by the analysis of \bianchiviih\ cosmologies,
we consider a deterministic global template and pose the analysis in
harmonic space (where \bianchiviih\ contributions can be computed
accurately and rotated efficiently).  The extension to alternative
settings, such as non-trivial topologies 
\citep{niarchou:2003,cornish:2004,dineen:2005,kunz:2005,bielewicz:2009},
is also highlighted.
Firstly, we describe a generic Bayesian analysis of anisotropic
cosmologies, before restricting this to the specific settings of full-
and partial-sky \cmb\ observations.  Finally, we perform simulations
to validate the implementation of the methodology presented.

\subsection{Bayesian analysis}
\label{sec:bayesian_analysis:bayesian_analysis}

By Bayes' theorem we may write the posterior distribution of the
parameters $\param$ of our model of interest $\fitmodelsel$, given
data $\fitdata$, as
\begin{equation}
 \prob(\param \given \fitdata, \fitmodelsel) 
 = \frac{\prob(\fitdata \given \param, \fitmodelsel) \, 
 \prob(\param \given \fitmodelsel)}{\prob(\fitdata \given \fitmodelsel)}
 \propto \prob(\fitdata \given \param, \fitmodelsel) \, 
 \prob(\param \given \fitmodelsel)
 \spcend ,
\end{equation}
where $\prob(\fitdata \given \param, \fitmodelsel)$ is the likelihood,
$\prob(\param \given \fitmodelsel)$ is the prior distribution of the
parameters of the model, and $\prob(\fitdata \given \fitmodelsel)$ is
the Bayesian evidence, which normalises the posterior distribution.
The (unnormalised) posterior distribution encodes our inferred
knowledge of the parameters of the model, given the observational
data.

We consider both models that include a \bianchiviih\ contribution in 
addition to a stochastic \cmb\ component and those that do not. We 
consider open and flat cosmologies where the cosmological parameters 
are given by $\cosmoparam=(A_s,\, n_s,\, \tau,\, \Denmatb 
\hubsmall^2,\, \Denmatc \hubsmall^2,\, \Denlambda,\, \Dencurvature)$, 
where $A_S$ is the amplitude of the primordial power
spectrum, $n_s$ is the scalar spectral index, $\tau$ is the optical
depth of reionisation, $\Denmatb \hubsmall^2$ is the physical baryon
density, $\Denmatc \hubsmall^2$ is the physical cold dark matter
density, $\Denlambda$ is the dark energy density, $\Dencurvature$ is
the curvature density and $\hub = 100\, \hubsmall$; for the flat case
$\Dencurvature=0$ and we recover the standard six parameter model. 
For the models including a \bianchiviih\ contribution, we consider 
the physical case where the Bianchi parameters \bparam\ and 
cosmological parameters \cosmoparam\ are coupled (\ie\ where the 
cosmological density parameters shared by \bparam\ and \cosmoparam\ 
are set equal) and also the non-physical case where they are not 
(the latter case is considered to make comparisons with previous work).
The \bianchiviih\ parameters \bparam\ are described in
\sectn{\ref{sec:bianchi_cosmologies}}.

For the case where a \bianchiviih\ template is embedded in a
stochastic \cmb\ background described by its power spectrum
$C_\el(\cosmoparam)$, the likelihood is given by
\begin{equation}
  \label{eqn:bianchi_likelihood}
  \prob( \fitdata  \given \bparam, \cosmoparam) \propto
  \frac{1}{\sqrt{|\mathbf{X}(\cosmoparam) |} } \exp{\Bigl[-\chi^2(\cosmoparam, \bparam) / 2\Bigr]}
  \spcend, 
\end{equation}
where
\begin{equation}
  \label{eqn:bianchi_chi2}
  \chi^2(\cosmoparam, \bparam) = 
  \bigl[\fitdata-\fittmpl(\bparam)\bigr]^\dagger \,
  \mathbf{X}^{-1}(\cosmoparam) \,
  \bigl[\fitdata-\fittmpl(\bparam)\bigr]
  \spcend ,
\end{equation}
and where $\fittmpl(\bparam)$ is the deterministic \bianchiviih\
template and $\mathbf{X}(\cosmoparam)$ is the covariance matrix of the
stochastic \cmb\ component (and any noise component included in the
model).  \Eqn{\ref{eqn:bianchi_likelihood}} and
\eqn{\ref{eqn:bianchi_chi2}} are written in a generic manner, where
the likelihood may be given either in real or harmonic space.  Since
\bianchiviih\ templates can be computed accurately and rotated
efficiently in harmonic space, we specialise to a harmonic space
representation, where the data and Bianchi template are given by their
spherical harmonic coefficients: $\fitdata = \{ \fitdataalm \}$ and
$\fittmpl(\bparam) = \{ \fittmplalm(\param) \}$, respectively,
considered up to the harmonic band-limit $\elmax$.  The covariance
matrix $\mathbf{X}(\cosmoparam)$ is then also specified in harmonic
space but differs depending on whether the full- or partial-sky
setting is considered (we consider each setting in turn in subsequent
subsections).  For the full-sky setting the covariance matrix is
diagonal, whereas this is not the case for the partial-sky setting
(assuming isotropic noise in both settings).

We have so far considered the setting where a subdominant,
deterministic template is embedded in a stochastic \cmb\ background.
However, it is straightforward to extend the analysis to the setting
where an anisotropic cosmology does not induce an embedded
deterministic template, but rather alters the covariance structure of
the stochastic \cmb\ component; this is the case for non-trivial
topologies (\eg\
\citealt{niarchou:2003,cornish:2004,dineen:2005,kunz:2005,bielewicz:2009}).
This setting can be handled in the current framework simply by setting
the template component to zero and by replacing the covariance matrix
$\mathbf{X}(\cosmoparam)$ with the appropriate covariance matrix for
the anisotropic cosmology.  The coordinate orientation is again
arbitrary in this setting, hence the covariance matrix, or similarly
the data, must be varied over all three-dimensional orientations as
specified by the Euler angles.  In this setting the $\chi^2$ of the
likelihood is written
\begin{equation}
\label{eqn:bianchi_chi2_alternative}
  \chi^2(\anisoparam) = 
  \fitdata^\dagger(\euls) \, \mathbf{X}^{-1}(\anisoparam) \, \fitdata(\euls)
  \spcend ,
\end{equation}
where $\anisoparam$ is the full set of parameters of the anisotropic
cosmology and $\fitdata(\euls)$ denotes the data rotated by Euler
angles $(\euls)$.  Since we focus on Bianchi cosmologies here, we
consider only the likelihood with $\chi^2$ given by
\eqn{\ref{eqn:bianchi_chi2}} in the remainder of this article.

To determine whether the inclusion of a \bianchiviih\ component better
describes the data the Bayesian evidence is examined, given by
\begin{equation}
  \evidence = 
  \prob(\fitdata \given \fitmodelsel) =
  \int \dx \param \,
  \prob(\fitdata \given \param, \fitmodelsel) \,
  \prob(\param \given \fitmodelsel) 
  \spcend .
\end{equation}
Using the Bayesian evidence to distinguish between models naturally
incorporates Occam's razor, trading off model simplicity and accuracy.
The odds ratio between two models $\fitmodelsel_1$ and
$\fitmodelsel_2$ may be written in terms of the Bayesian evidences of
the models ($\evidence_1$ and $\evidence_2$ respectively) by
\begin{equation}
  \frac{\prob(\fitmodelsel_1 \given \fitdata)}{\prob(\fitmodelsel_2
    \given \fitdata)} = 
  \frac{\evidence_1}{\evidence_2} \,
  \frac{\prob(\fitmodelsel_1)}{\prob(\fitmodelsel_2)}
  \spcend .
\end{equation}
In the absence of any prior information about a preferred model, \ie\
when $\prob(\fitmodelsel_1) = \prob(\fitmodelsel_2)$, the Bayes factor
given by the ratio of Bayesian evidences $E_1/E_2$ is thus identical
to the ratio of the model probabilities given the data.  Consequently,
the Bayes factor may be used to distinguish models in this setting.

The Jeffreys scale \citep{jeffreys:1961} is often used as a
rule-of-thumb when comparing models via their Bayes factor. The
log-Bayes factor $\Delta {\rm ln} E = {\rm ln} (E_1/E_2)$ 
represents the degree by which model
$\fitmodelsel_1$ is favoured over model $\fitmodelsel_2$, assuming the 
models are equally likely {\em a priori}. On the Jeffreys scale log-Bayes
factors are given the following interpretation: $0 \leq
\Delta {\rm ln} E < 1$ is regarded as inconclusive; $1 \leq \Delta
{\rm ln} E < 2.5$ as significant; $2.5 \leq \Delta {\rm ln} E < 5$ as
strong; and $\Delta {\rm ln} E \geq 5 $ as conclusive (without loss of
generality we have assumed $E_1 \geq E_2$).  For reference, a
log-Bayes factor of 2.5 corresponds to odds of approximately 1 in 12,
while a factor of 5 corresponds to odds of approximately 1 in 150.

\subsection{Full-sky analysis}
\label{sec:bayesian_analysis:full_sky}

In the full-sky setting the covariance matrix is given by
\mbox{$\mathbf{X}(\cosmoparam) = \mathbf{C}(\cosmoparam)$}, where
$\mathbf{C}(\cosmoparam)$ is the diagonal \cmb\ covariance matrix with
entries given by the power spectrum $C_\el(\cosmoparam)$ on its
diagonal.  A Bayesian analysis in this setting was first considered
by \cite{bridges:2006b}.  To expose the detail of the analysis we make
the likelihood explicit in this setting. The likelihood, in terms of
the spherical harmonic coefficients of the data $\fitdataalm$ and
\bianchiviih\ template $\fittmplalm(\bparam)$, is given by
\begin{align}
  \prob( \{ \fitdataalm \}  \given \bparam, \cosmoparam) \propto
  &\prod_{\el=0}^\elmax \frac{1}{\sqrt{C_\el(\cosmoparam)}}
    \exp{ \Biggl\{ -\frac{\left[\fitdataalz-\fittmplalz(\bparam)\right]^2}{C_\el(\cosmoparam)}\Biggr\}} \nonumber\\
  &\times
  \prod_{\m=1}^\el \frac{2}{C_\el(\cosmoparam)} 
    \exp{\Biggl\{-\frac{\left|\fitdataalm-\fittmplalm(\bparam)\right|^2}{C_\el(\cosmoparam)}\Biggr\}}
  \spcend.
\end{align}
In practice, it is more convenient numerically to work with the
log-likelihood function, given by
\begin{align}
  {\rm ln} [ \prob( & \{ \fitdataalm \}  \given \bparam, \cosmoparam) ]
  \propto
  \sum_{\el=0}^\elmax \Biggl\{  
  (2\el+1) {\rm ln}[C_\el(\cosmoparam)] 
  \nonumber \\
   &+ \frac{[\fitdataalz-\fittmplalz(\bparam)]^2}{C_\el(\cosmoparam)}
   + \frac{2}{C_\el(\cosmoparam)} \sum_{\m=1}^\el \left|\fitdataalm-\fittmplalm(\bparam)\right|^2
  \Biggr \}
  \spcend .
  \label{eqn:fit_loglikelihood:2}
\end{align}
In the case of a zero Bianchi component $\fittmpl(\bparam)=0$,
\eqn{\ref{eqn:fit_loglikelihood:2}} reduces to the log-likelihood
function used commonly to compute parameter estimates from the power
spectrum estimated from \cmb\ data (\eg\ \citealt{verde:2003}).

\subsection{Partial-sky analysis}
\label{sec:bayesian_analysis:partial_sky}

In the partial-sky setting a mask is applied in real space to excise
contaminated regions of the data (due to point-source or Galactic
contamination, for example).  See, for example, the mask applied to
simulated \cmb\ data in \fig{\ref{fig:sim_maps_partialsky:cmb}}.  

Various approaches can be adopted to deal with partial-sky
  data.  The standard approach is to multiply the map by a binary
  mask.  In this setting, the application of a mask induces coupling
  of harmonic space modes in the resultant data, which can be viewed
  as a convolution of the original covariance matrix and the mask
  transfer function in harmonic space (the so-called coupling matrix).
  This coupling can then be taken into account when making a
  comparison with theory \citep[\eg][]{hivon:2002,hinshaw:2003}.
  However, we take a different approach here, which avoids the need to
  compute the coupling matrix explicitly (thereby avoiding the
  computation of Wigner 3-$j$ symbols).

We effectively consider full-sky data but add \textit{masking
    noise} to marginalise over the pixel values of the data in the masked
  region.  Gaussian masking noise $\mnoise$ is added to the data
$\fitdatasky$ in real space:
\begin{equation}
\label{eqn:effective_data}
s(\sa) = \fitdatasky(\sa) + \mnoise(\sa)
\spcend,
\end{equation}
where $\sa$ denotes angular coordinates on the sphere.  
The masking noise is chosen to be zero-mean and large in the masked
region of the data (\ie\ zeros of the mask), and zero elsewhere.
Consequently, the masking noise is anisotropic over the sky but may be
chosen to be uncorrelated, and may thus be defined by its covariance
\begin{equation}
\label{eqn:mnoise_covariance_real}
\langle 
\mnoise(\sa_i) \, \mnoise^\cconj(\sa_j)
\rangle
= \kron{i}{j}\,
\sigma_\mnoise^2(\sa_i)
\spcend ,
\end{equation}
where $\kron{i}{j}$ is Kronecker delta symbol, $\sa_i$ denotes the
angular coordinate of pixel $i$, and the variance of the noise for
pixel $i$ is given by
\begin{equation}
\label{eqn:mnoise_variance}
\sigma_\mnoise^2(\sa_i)
= 
\begin{cases}
\Sigma_\mnoise^2, & \sa_i \in \mathbb{M} \\
0, & \sa_i \in \sphere \backslash \mathbb{M} \\
\end{cases}
\spcend ,
\end{equation}
where $\Sigma_\mnoise^2$ is a constant masking noise variance.  We
adopt the \healpix\ \citep{gorski:2005} equal-area pixelisation of the
sphere, where the area of each pixel is $\pixarea$, in order
to avoid dealing with pixels of differing areas.  Here $\mathbb{M}$
denotes the masked region of the sky $\sphere$ and $\sphere \backslash
\mathbb{M}$ denotes the remaining region. By synthetically adding
masking noise that is much larger than the original data in the masked
region of the sky, we effectively marginalise over the pixel values of the
data in this region.  Consequently, only the pixel values of the data
outside of the masked region have a large influence on the final
analysis.  We refer to a particular masking noise realisation as a
\textit{noisy mask}.  An example noisy mask is shown in
\fig{\ref{fig:sim_maps_partialsky:noisy_mask}}.

The noisy mask introduces coupling in harmonic space that must be accounted for in the analysis.  The effective
signal $s$ to be analysed now includes the original data $\fitdatasky$
and the noisy mask $\mnoise$; thus its covariance is the sum of the
covariance of the original data and the noisy mask.  The covariance of
the noisy mask is given in harmonic space by
\begin{equation}
\label{eqn:mnoise_covariance_harmonic}
\mathbf{M}_{\el\m}^{\el\p\m\p} 
= 
\langle
\shc{\mnoise}{\el}{\m} \,
\shcc{\mnoise}{\el\p}{\m\p}
\rangle 
\simeq
\sum_{\sa_i} 
\sigma_\mnoise^2(\sa_i)
\shfargc{\el}{\m}{\sa_i} \,
\shfarg{\el\p}{\m\p}{\sa_i} \,
\pixarea^2
\spcend ,
\end{equation}
where we have applied \eqn{\ref{eqn:mnoise_covariance_real}} and again
consider signals and their covariance structure in harmonic space
(since the analysis of \bianchiviih\ cosmologies is most accurately and
efficiently performed in harmonic space).
In \eqn{\ref{eqn:mnoise_covariance_harmonic}} we approximate the
spherical harmonic coefficients of the mask using discrete quadrature:
\begin{equation}
\shc{\mnoise}{\el}{\m} 
= \int_\sphere \dmu{\sa} \,
\mnoise(\sa) \,
\shfargc{\el}{\m}{\sa} \,
\simeq
\sum_{\sa_i} 
\pixarea \,
\mnoise(\sa_i) \,
\shfargc{\el}{\m}{\sa_i}
\spcend ;
\end{equation}
note that this is necessarily an approximation since the noisy mask is
not band-limited and we adopt the {\tt HEALPix} \citep{gorski:2005}
sampling scheme, which does not admit exact quadrature.  Since the underlying data $\fitdatasky$ and noisy mask $\mnoise$ are independent, the
covariance of the effective data $s$ is given by
\mbox{$\mathbf{X}(\cosmoparam) = \mathbf{C}(\cosmoparam) +
  \mathbf{M}$}, where $\mathbf{C}(\cosmoparam)$ is again the diagonal
\cmb\ covariance matrix and $\mathbf{M}$ is the non-diagonal noisy
mask covariance matrix given by \eqn{\ref{eqn:mnoise_covariance_harmonic}}.

To summarise, the partial-sky analysis proceeds as follows.  Firstly a
zero-mean, Gaussian noisy mask realisation is constructed via
\eqn{\ref{eqn:mnoise_covariance_real}} and
\eqn{\ref{eqn:mnoise_variance}}, and its covariance structure is
estimated by \eqn{\ref{eqn:mnoise_covariance_harmonic}}. The effective
signal under analysis $s$ is constructed by summing the original data
and the noisy mask realisation by \eqn{\ref{eqn:effective_data}}.  The
Bayesian analysis described in
\sectn{\ref{sec:bayesian_analysis:bayesian_analysis}} is performed,
where the data under analysis is replaced with the effective signal
$s$, with covariance matrix given by \mbox{$\mathbf{X}(\cosmoparam) =
  \mathbf{C}(\cosmoparam) + \mathbf{M}$}.\footnote{In practice the
  masked region of the data to be excised will include unknown
  contamination, rather than a \cmb\ contribution only.  Thus, the
  original data $\fitdatasky$ are masked before the procedure outlined
  here is performed.  In this setting the covariance matrix
  $\mathbf{X}(\cosmoparam) = \mathbf{C}(\cosmoparam) + \mathbf{M}$ is
  necessarily an approximation, since the assumption that
    there is a CMB component in the masked region is no longer valid.
  However, provided the noise variance $\Sigma_\mnoise^2$ is chosen to
  be considerably larger than the expected \cmb\ contribution, the
  approximation is very accurate.}

\subsection{Implementation and validation}
\label{sec:bayesian_analysis:implementation}

\begin{figure}
\centering
\subfigure[\cmb\ component]{\includegraphics[width=0.72\columnwidth]{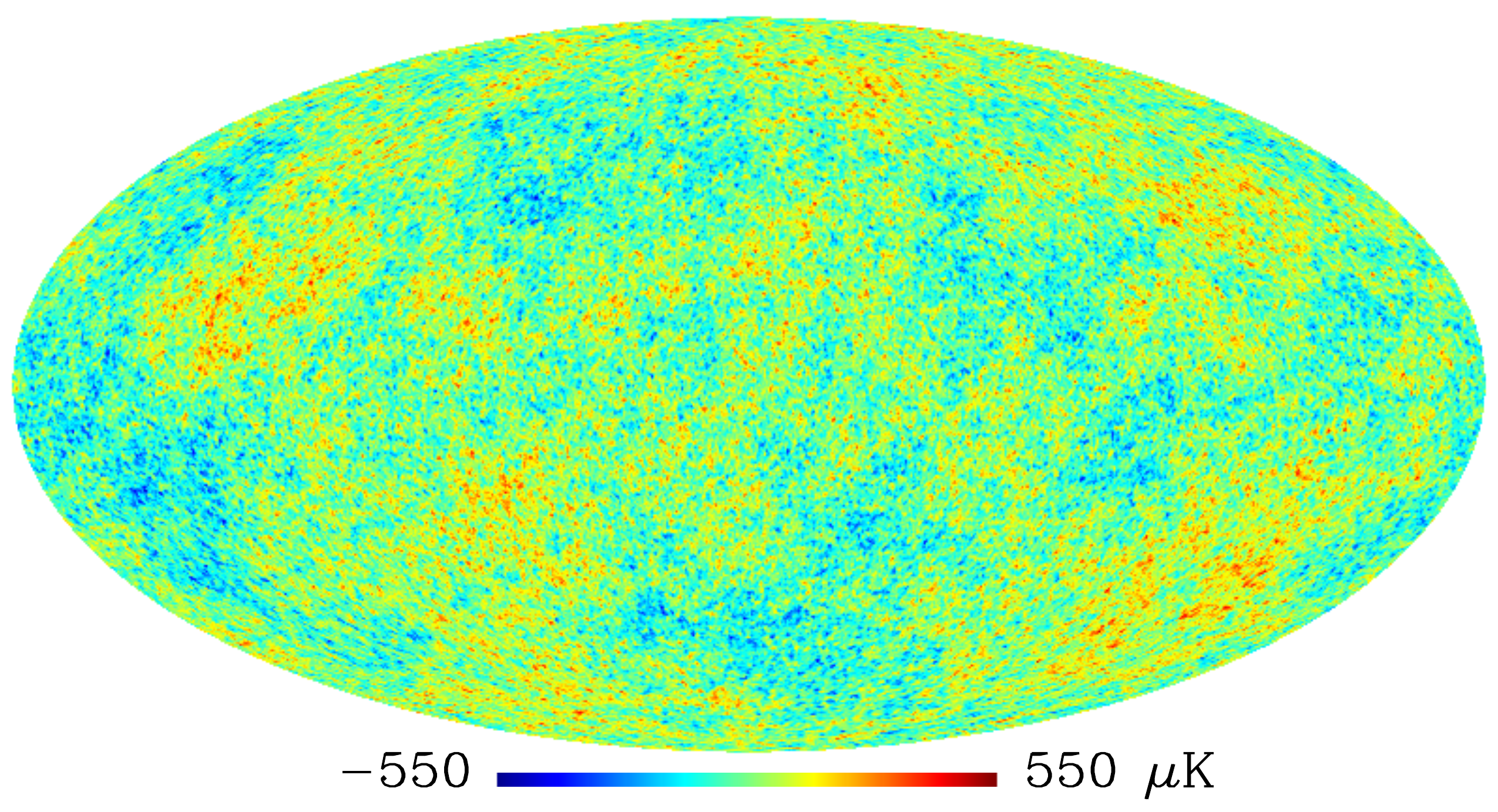}}
\subfigure[Underlying Bianchi component \label{fig:sim_maps_fullsky:embedded}]{\includegraphics[width=0.72\columnwidth]{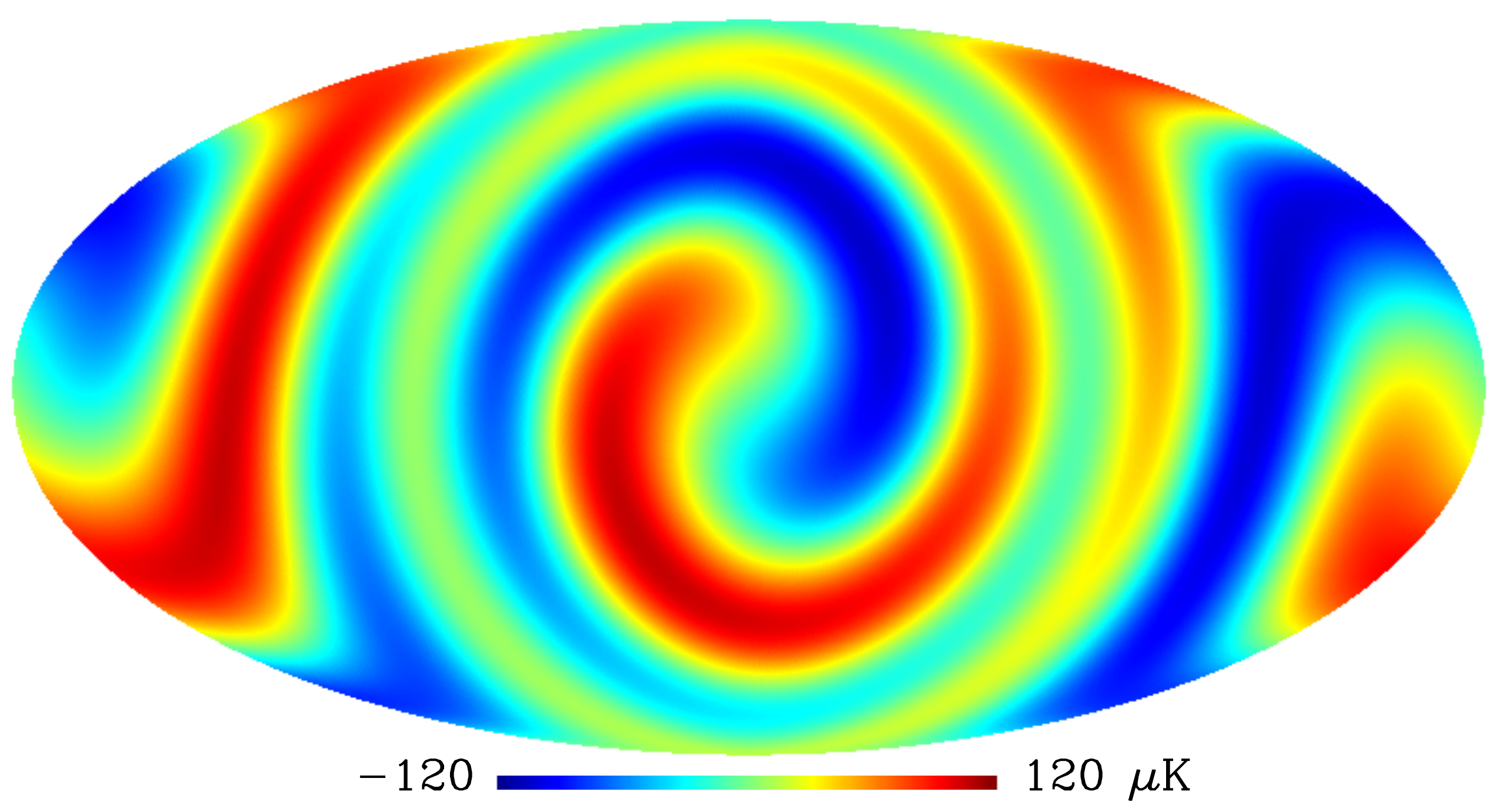}}
\subfigure[\cmb\ plus Bianchi components]{\includegraphics[width=0.72\columnwidth]{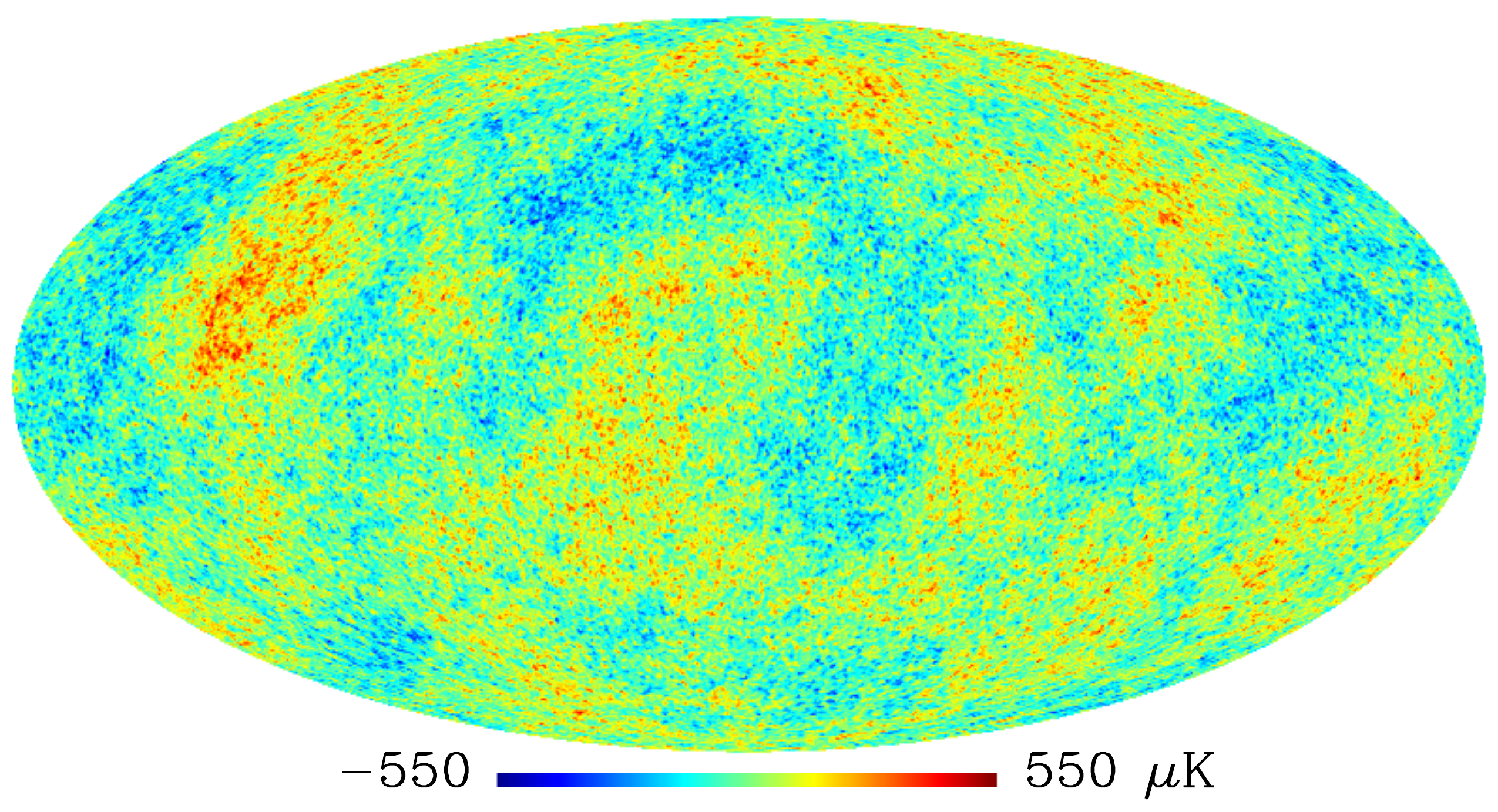}}
\subfigure[Recovered Bianchi component \label{fig:sim_maps_fullsky:recovered}]{\includegraphics[width=0.72\columnwidth]{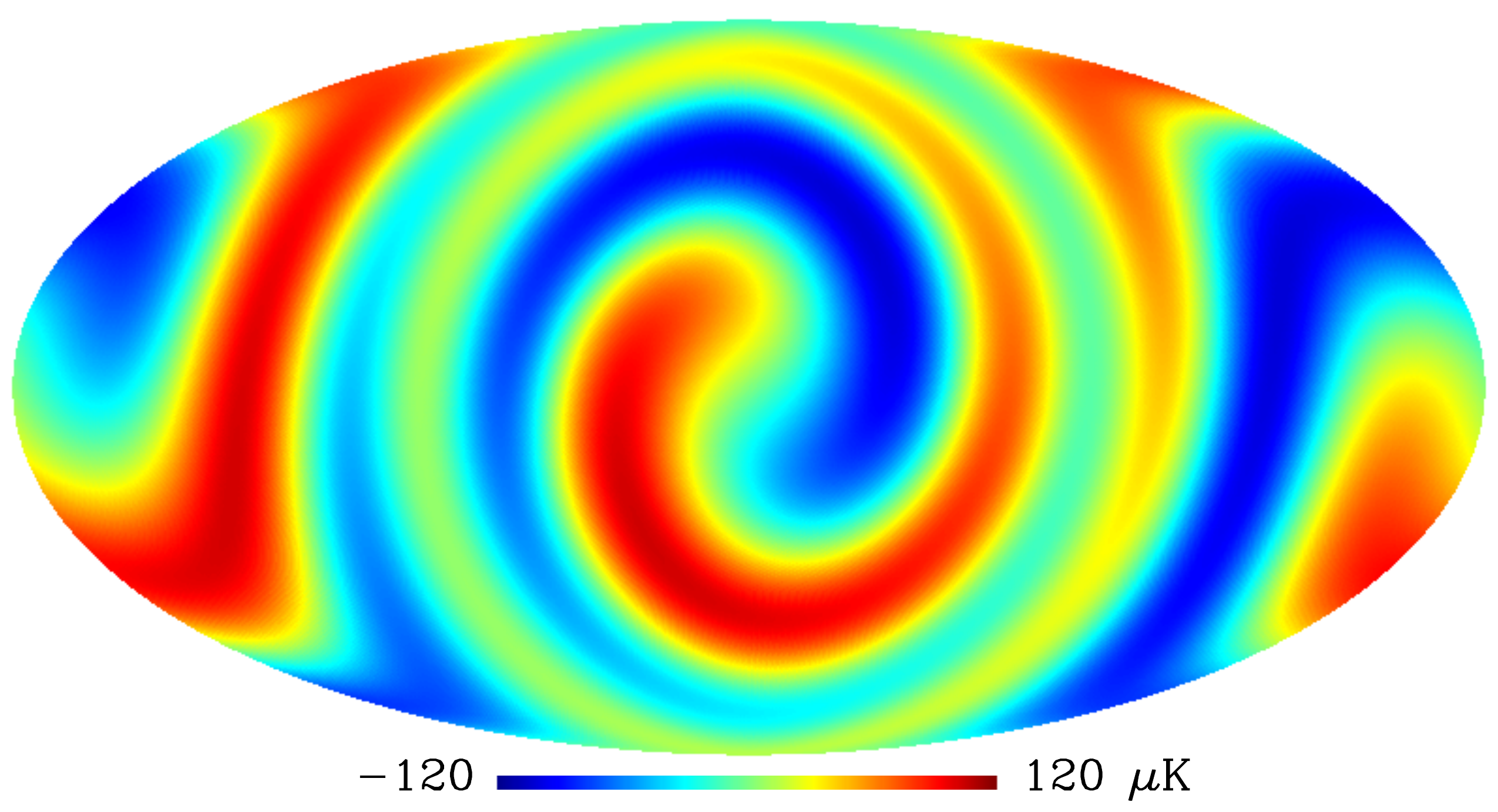}}
\caption{Full-sky simulation at $\ell_{\rm max}=512$. A \bianchiviih\
  component with a large amplitude is simulated (panel b) and embedded
  in a standard stochastic \cmb\ map (panel a), yielding a resultant
  \cmb\ plus Bianchi map (panel c) to which a beam and noise are
  applied.  The full-sky Bayesian analysis described in
  \sectn{\ref{sec:bayesian_analysis:full_sky}} is applied to recover a
  MAP estimated Bianchi component (panel d).  The recovered Bianchi
  component (panel d) accurately estimates the embedded component
  (panel b).}
\label{fig:sim_maps_fullsky}
\end{figure}

\begin{figure*}
\centering
\includegraphics[width=0.90\textwidth]{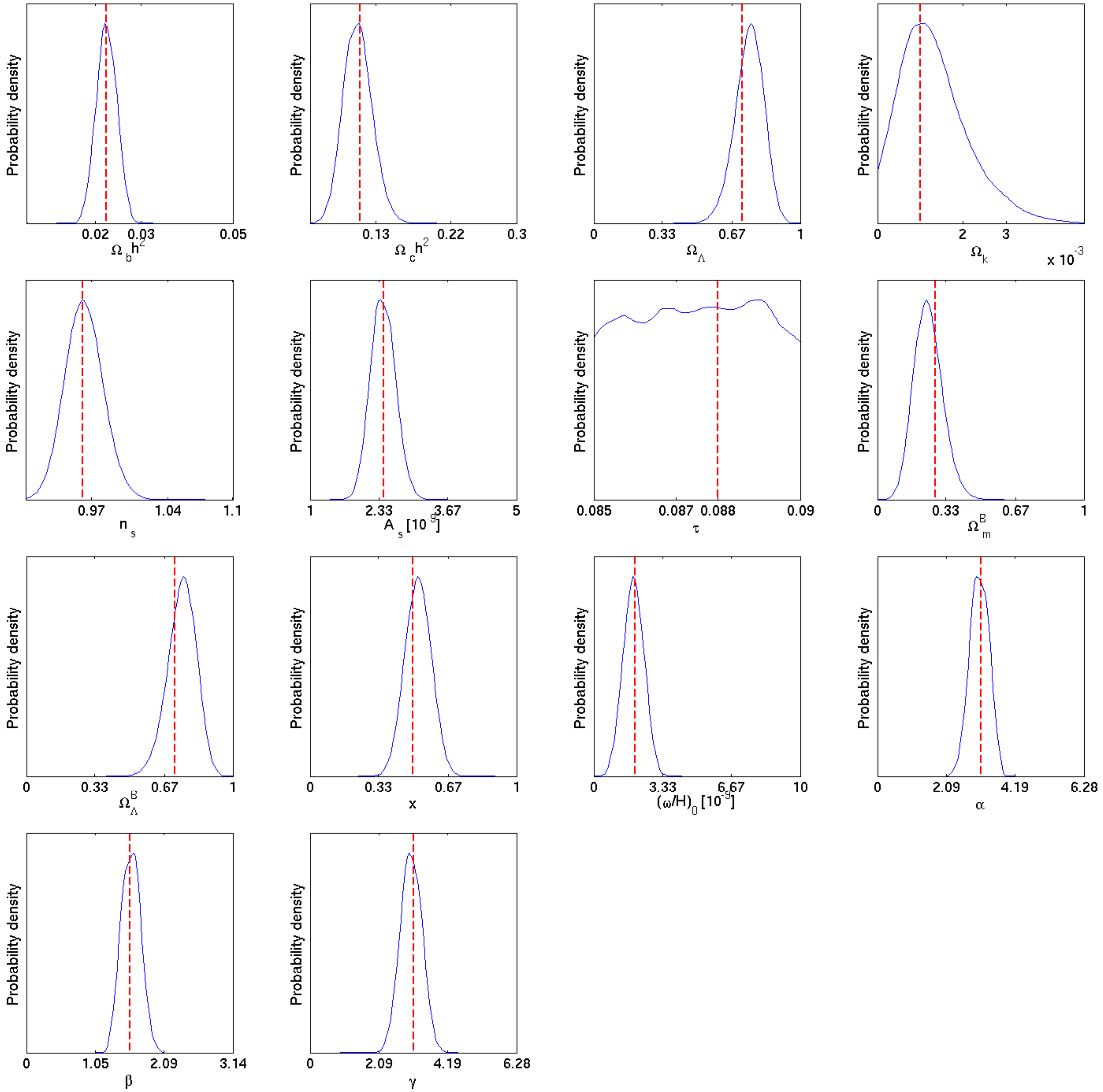}
\caption{Posterior distributions corresponding to the full-sky
  simulation at $\elmax=512$ shown in
  \fig{\ref{fig:sim_maps_fullsky}}.  The underlying parameter values
  of the simulation are indicated by dashed vertical lines.  Both
  cosmological and \bianchiviih\ parameters are estimated accurately,
  except for the optical depth of reionisation, $\tau$, which is not
  unexpected when using \cmb\ temperature data alone. For model
  comparison purposes an unconstrained $\tau$ does not pose any
  great concern.}
\label{fig:sims_posterior_fullsky}
\end{figure*}

\begin{figure}
\centering
\subfigure[Masked \cmb\ component \label{fig:sim_maps_partialsky:cmb}]{\includegraphics[width=0.72\columnwidth]{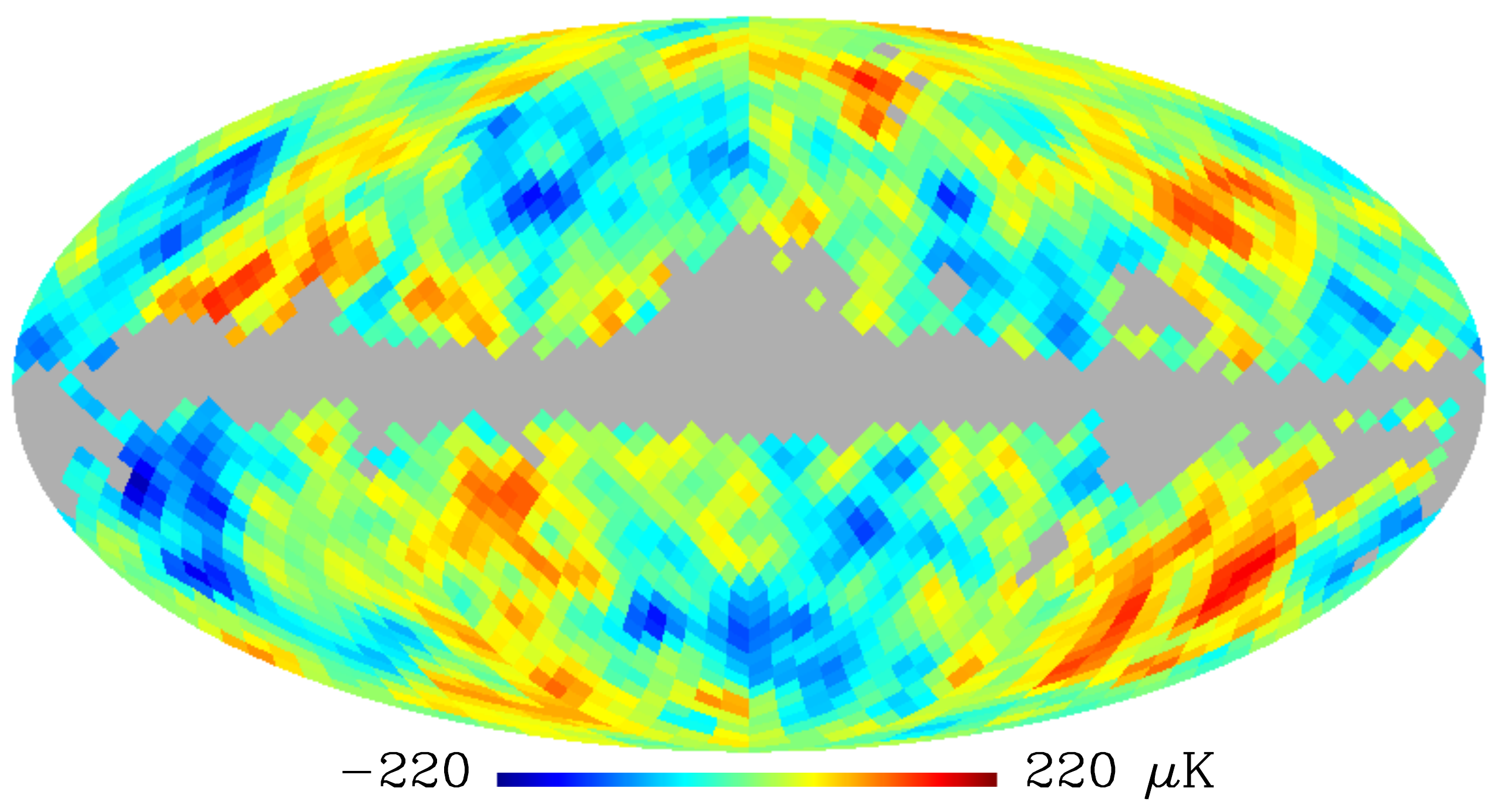}}
\subfigure[Underlying Bianchi component \label{fig:sim_maps_partialsky:embedded}]{\includegraphics[width=0.72\columnwidth]{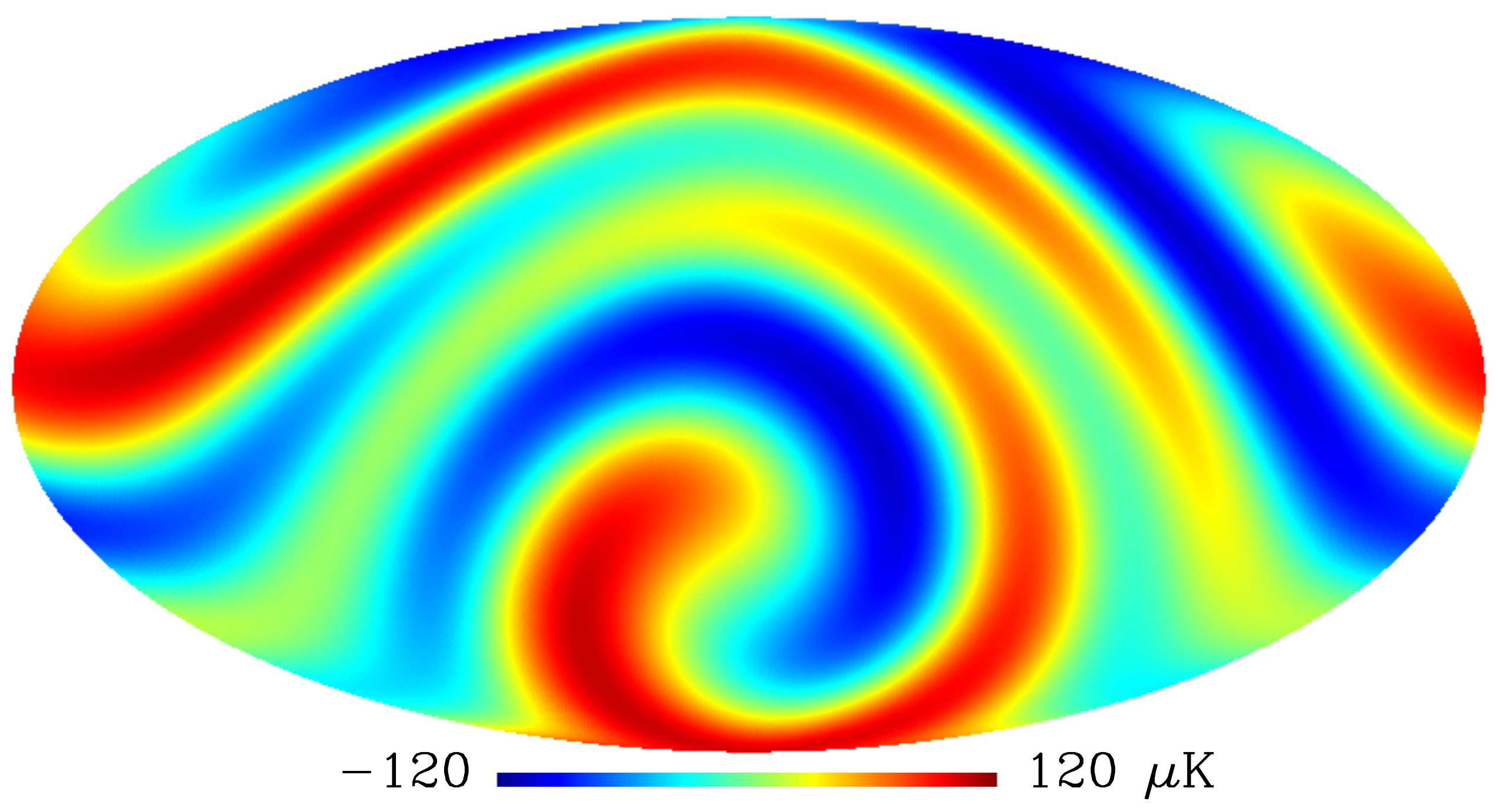}}
\subfigure[Masked \cmb\ plus Bianchi components]{\includegraphics[width=0.72\columnwidth]{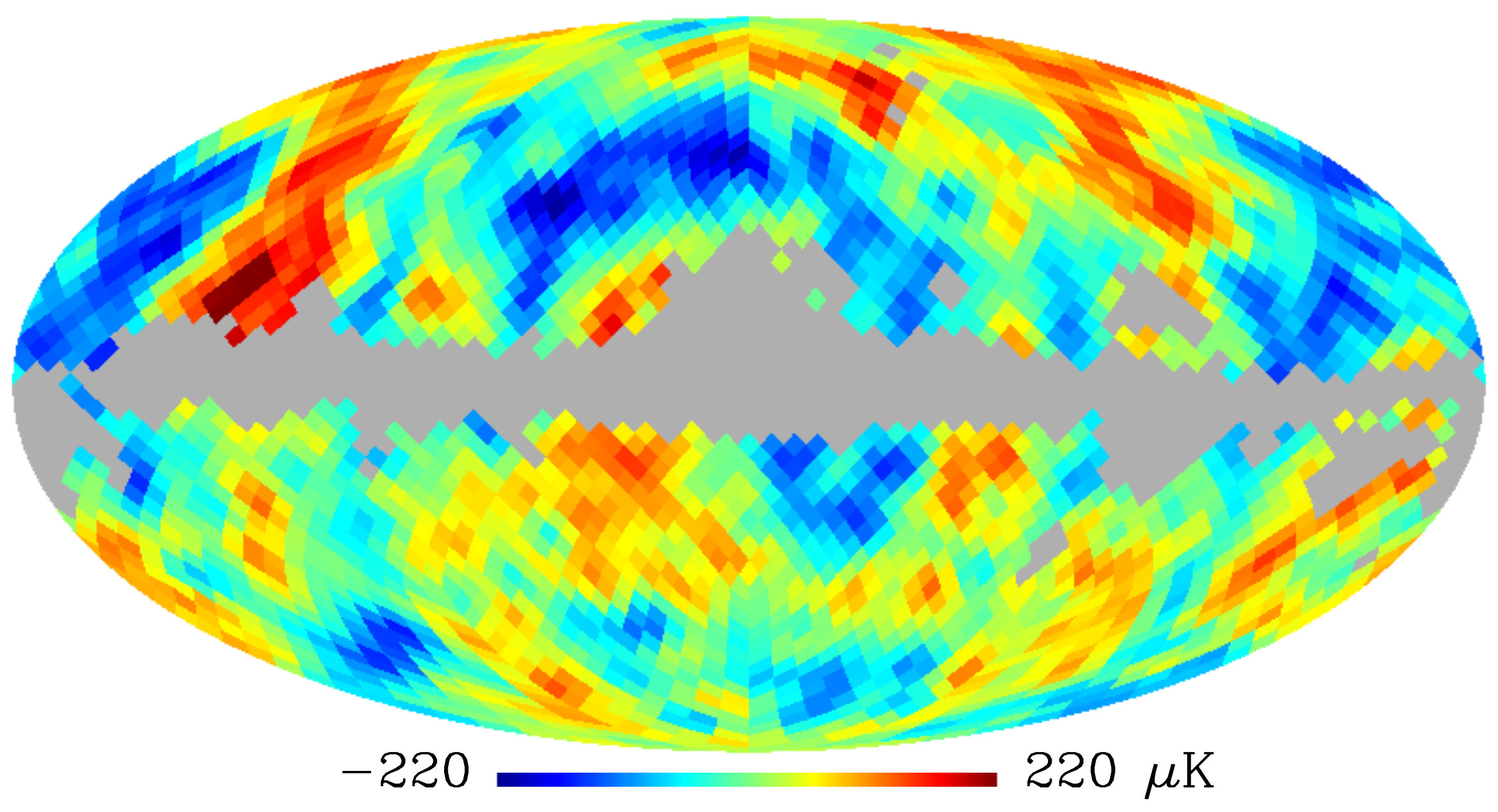}}
\subfigure[Noisy mask realisation \label{fig:sim_maps_partialsky:noisy_mask}]{\includegraphics[width=0.72\columnwidth]{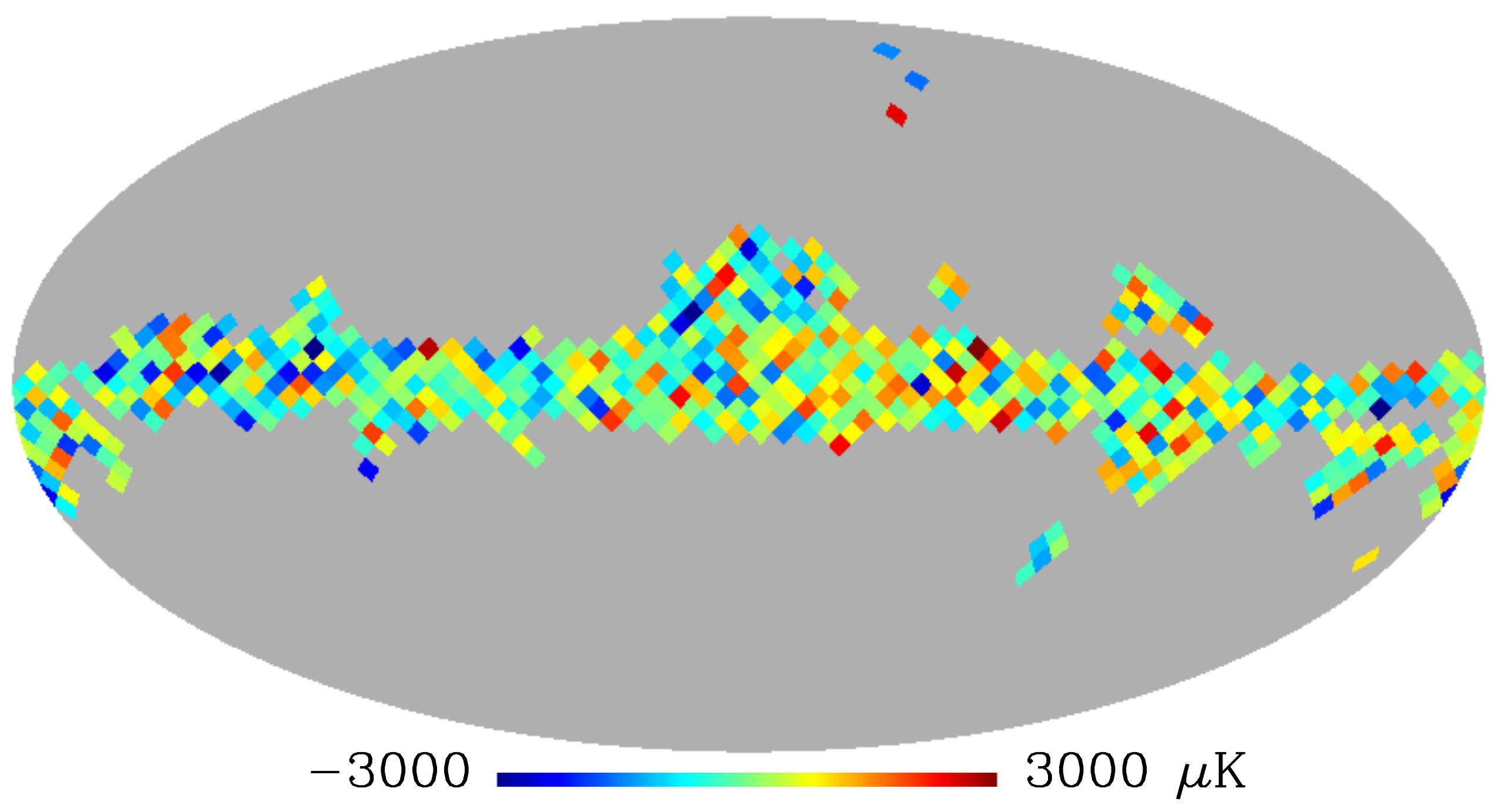}}
\subfigure[Recovered Bianchi component \label{fig:sim_maps_partialsky:recovered}]{\includegraphics[width=0.72\columnwidth]{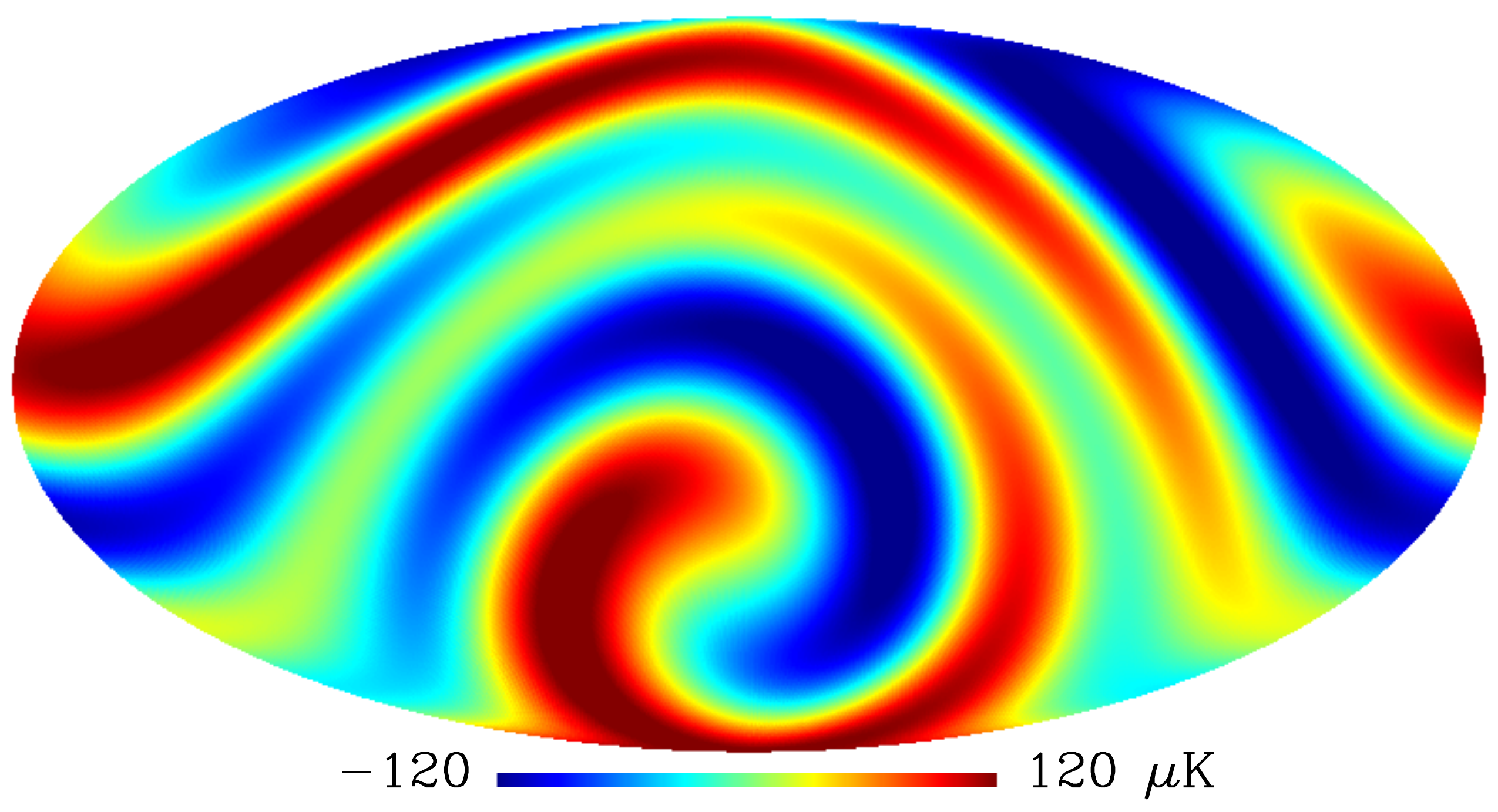}}
\caption{Partial-sky simulation at $\elmax=32$.  A \bianchiviih\
  component with a large amplitude is simulated (panel b) and embedded
  in a standard stochastic \cmb\ map (panel a), yielding a resultant
  \cmb\ plus Bianchi map (panel c) to which a mask, beam and noise are
  applied (the Bianchi component is rotated such that the majority of
  its structure does not lie in the masked region).  A noisy mask
  (panel d) is added to the \cmb\ plus Bianchi map and the partial-sky
  Bayesian analysis described in
  \sectn{\ref{sec:bayesian_analysis:partial_sky}} is applied to
  recover a MAP estimated Bianchi component (panel e).  The recovered
  Bianchi component (panel e) accurately estimates the embedded
  component (panel b). }
\label{fig:sim_maps_partialsky}
\end{figure}

\begin{figure*}
\centering
\includegraphics[width=0.90\textwidth]{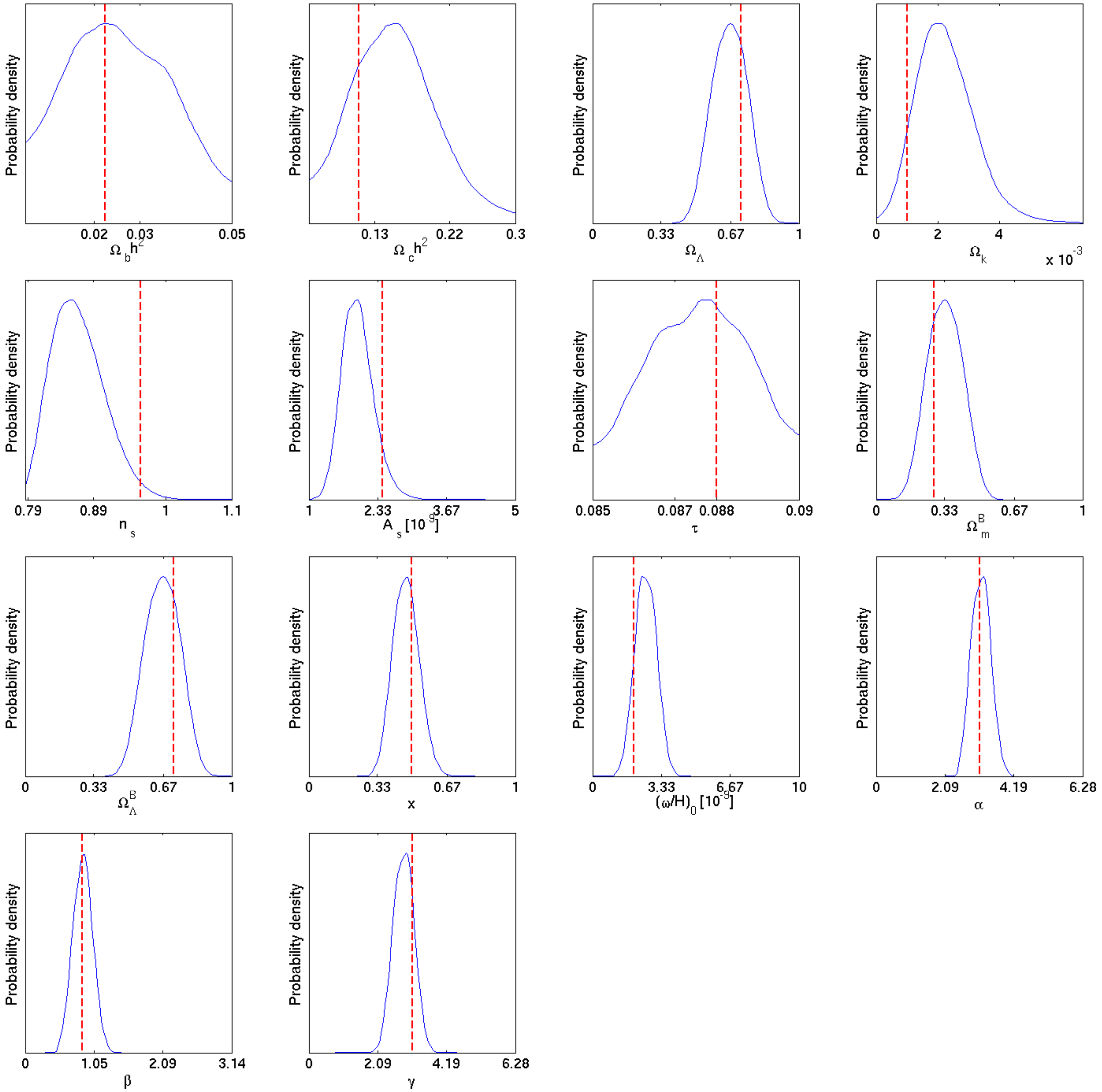}
\caption{Posterior distributions corresponding to the partial-sky
  simulation at $\elmax=32$ shown in
  \fig{\ref{fig:sim_maps_partialsky}}. The underlying parameter values
  of the simulation are indicated by dashed vertical lines.  Both
  cosmological and \bianchiviih\ parameters are estimated reasonably
  well, however since less data is now available due to the lower
  \elmax\ and masking, the marginalised posterior distributions are
  not as accurate or well constrained as the full-sky setting, as is
  to be expected.  For model comparison purposes this does not pose
  any great concern.}
\label{fig:sims_posterior_partialsky}
\end{figure*}

We have developed the \anicosmo\footnote{The \anicosmo\ code will soon
  be made publicly available from
  \mbox{\url{http://www.jasonmcewen.org/}}.} code to perform a
Bayesian analysis of anisotropic cosmologies.  The Bayesian analysis
described in this section is implemented in \anicosmo, both in the
full- and partial-sky settings.  We adopt the {\tt
  MultiNest}\footnote{\url{http://www.mrao.cam.ac.uk/software/multinest/}}
code \citep{feroz:multinest1,feroz:multinest2} to explore the
posterior distribution and to compute Bayesian evidence values, both
via nested sampling
\citep{skilling:2004,feroz:multinest1,feroz:multinest2}.  The {\tt
  CAMB}\footnote{\url{http://camb.info/}} code \citep{lewis:2000} is
used to compute the \cmb\ power spectrum $C_\el(\cosmoparam)$ for
various cosmological parameters $\cosmoparam$.

At present the \anicosmo\ code is specialised to the study of
\bianchiviih\ cosmologies, where the likelihood is computed through
\eqn{\ref{eqn:bianchi_likelihood}} and \eqn{\ref{eqn:bianchi_chi2}}.
However, \anicosmo\ may be trivially extended to handle the
$\chi^2$ defined by \eqn{\ref{eqn:bianchi_chi2_alternative}} and to
study other anisotropic cosmologies, such as non-trivial topologies;
this will be the focus of future work.

In the remainder of this subsection we perform simulations to validate
the Bayesian analysis method described here and its implementation in
the \anicosmo\ code.  In both the full- and partial- sky settings, we
simulate a \cmb\ map and synthetically embed a simulated \bianchiviih\
component.  To approximate the finite-resolution, noisy measurements
of the \cmb\ made by the \wmap\ experiment (specifically those made in
the highest-resolution, 94 GHz W band; \citealt{bennett:2013}), a
Gaussian beam with full-width-half-maximum (FWHM) $13.2\arcmin$ and
isotropic noise with power spectrum $N_\el=0.02\,(\mu{\kelvin})^2$ are
applied to the simulations. To account for the beam and noise in the
Bayesian analysis we simply map $C_\el(\cosmoparam) \rightarrow
b_\el^2 \, C_\el(\cosmoparam) + N_\el$ when computing the \cmb\
covariance matrix and apply the beam to the \bianchiviih\ template by
mapping $\fittmplalm(\bparam) \rightarrow b_\el \,
\fittmplalm(\bparam)$, where $b_\el^2$ and $N_\el$ are the beam and
noise power spectra, respectively.  In these simulations we consider
the model where the Bianchi parameters \bparam\ and cosmological
parameters \cosmoparam\ are coupled.  Posterior distributions and
evidence values are then recovered by \anicosmo\ for models including
and excluding a coupled Bianchi component in addition to the standard
stochastic \cmb\ component. For the purpose of validating the
implementation, we consider a Bianchi template with a relatively large
amplitude of $\bvortinline=2 \times 10^{-9}$.

Full-sky simulations are shown in \fig{\ref{fig:sim_maps_fullsky}}.
Since the covariance matrix $\mathbf{X}(\cosmoparam)$ is diagonal in
this setting, the analysis may be run at high-resolution; hence, we
set $\elmax=512$ in this simulation.  The \textit{maximum-a-posteriori}
(MAP) estimated \bianchiviih\ component recovered from the peak of the
posterior distribution $\prob(\param \given \fitdata, \fitmodelsel)$
is shown in \fig{\ref{fig:sim_maps_fullsky:recovered}} and is clearly
an accurate estimate of the embedded component shown in
\fig{\ref{fig:sim_maps_fullsky:embedded}}.  The marginalised posterior
distributions recovered for each parameter are shown in
\fig{\ref{fig:sims_posterior_fullsky}}.  Both cosmological and
\bianchiviih\ parameters are estimated accurately, except for the
optical depth of reionisation $\tau$ which is not unexpected when
using \cmb\ temperature data alone. $\tau$ is also unconstrained when
considering the model absent of a Bianchi component, thus for model
comparison purposes an unconstrained $\tau$ does not pose any great concern.
Indeed, for this simulation the model including a Bianchi component is
favoured conclusively with a log-Bayes factor of $\Delta {\rm ln}
E\sim 50$.

Partial-sky simulations are shown in
\fig{\ref{fig:sim_maps_partialsky}}.  Since the covariance matrix
$\mathbf{X}(\cosmoparam)$ is non-diagonal in this setting, the
analysis is considerably more computationally demanding than the
full-sky setting; hence, we set $\elmax=32$ in this simulation.  The
MAP estimated \bianchiviih\ component is shown in
\fig{\ref{fig:sim_maps_partialsky:recovered}} and clearly is an
accurate estimate of the embedded component shown in
\fig{\ref{fig:sim_maps_partialsky:embedded}} (note that the embedded
Bianchi component was rotated such that the majority of its structure
does not lie in the masked region).  A noisy mask variance of
$\Sigma_\mnoise^2=100\,({\rm m\kelvin})^2$ was adopted since numerical
tests showed this effectively marginalised masked pixels in the data,
while increasing $\Sigma_\mnoise^2$ further had little impact. The
marginalised posterior distributions recovered for each parameter are
shown in \fig{\ref{fig:sims_posterior_partialsky}}.  Both cosmological
and \bianchiviih\ parameters are estimated reasonably accurately,
however since less data is now available due to the lower \elmax\ and
masking, the marginalised posterior distributions are not as well
constrained as the full-sky setting, as is to be
expected.\footnote{Notice that the optical depth $\tau$
    appears to be better constrained in the partial-sky setting than
    in the full-sky setting. This is most likely a chance result for
    the particular simulations considered and is not expected to hold
    in general.  Indeed, when considering the \wmap\ data (with the
    same $\elmax$) the recovered posterior distributions for the
    optical depth are very similar (see
    \fig{\ref{fig:posterior_opencoupledbianchi}} and
    \fig{\ref{fig:posterior_flatdecoupledbianchi}}).}

Reducing the maximum multipole considered has a large impact on the 
\lcdm\ parameters determined by the acoustic peak positions and heights, 
but, thanks to the effective low-$\ell$ band-limit of the \bianchiviih\ 
signal, affects the \bianchiviih\ parameters much more mildly. This is 
clearly evident from the comparative behaviour of the posteriors for
\bvortinline\ and, \eg, $\Denmatc \hubsmall^2$ or $\Denmatb \hubsmall^2$.
For model comparison purposes, this does not therefore pose any great 
concern.  Indeed, for this simulation the model including a Bianchi 
component is also favoured conclusively with a log-Bayes factor of 
$\Delta {\rm ln} E\sim 50$. Nevertheless, by reducing $\elmax$ we 
clearly discard a great deal of cosmologically interesting information. 
This can be recovered by calculating a high-$\ell$ likelihood 
assuming the \bianchiviih\ contribution is zero for $\ell > \elmax$, using a 
conjugate-gradient-based implementation 
\citep{smith:2007,bennett:2013} of the 
optimal, unbiased power-spectrum estimator \citep{tegmark:1997} in the presence of a mask.
The two likelihoods can then be multiplied, assuming that the high- 
and low-$\ell$ data are independent. The implementation of such an
algorithm for the simulations used here is unnecessary, as they are
intended solely to validate the \anicosmo\ code. When 
considering the \wmap\ data, we will use the purpose-built \wmap\
likelihood code\footnote{Available for download from 
\url{http://lambda.gsfc.nasa.gov/product/map/dr5/likelihood_get.cfm}.} 
\citep{bennett:2013}, which implements the scheme described 
above, to calculate the high-$\ell$ likelihood and hence maximise the
impact of the \wmap\ data.

The rationale for the simulations performed here is to validate the
implementation of the \anicosmo\ code, which has been demonstrated
effectively in both the full- and partial-sky settings.  Furthermore,
although the details are not presented here, additional simulations
were performed to validate the partial-sky setting.  Two \cmb\ maps
were simulated with different cosmologies.  A single hybrid map was
then constructed with the Northern hemisphere given by the \cmb\
simulated from the first cosmology and with the Southern hemisphere
given by the \cmb\ simulated from the second cosmology. \anicosmo\ was
applied to recover cosmological parameters from this hybrid map, once
masking the Northern hemisphere and once masking the Southern
hemisphere.  In each of these tests the correct cosmology of the
unmasked hemisphere was recovered, demonstrating that the partial-sky
Bayesian analysis outlined in
\sectn{\ref{sec:bayesian_analysis:partial_sky}} effectively
marginalises the masked pixels of the data.  This test and the
simulations described previously provide a strong validation of the
\anicosmo\ code in both the full- and partial-sky settings.

\section{Analysis of \wmap\ observations}
\label{sec:wmap_analysis}

We analyse \wmap\ 9-year data for evidence of \bianchiviih\
cosmologies, performing both the full- and partial-sky Bayesian
analyses described in the preceding section.  Firstly, we describe
the specific full- and partial-sky \wmap\ data used and the
cosmological models considered.  We then present the results of our
Bayesian analysis of \bianchiviih\ models and place robust constraints
on the vorticity of the Universe.

\subsection{Data}
\label{sec:wmap_analysis:data}

Previous searches for \bianchiviih\ components embedded in \wmap\ data 
(\eg\ \citealt{bridges:2006b}) used full-sky data in the form 
of the ILC map to estimate the parameters of the \bianchiviih\ model. While 
this greatly simplifies the form of the covariance matrix employed in the 
likelihood, the ILC map was not recommended for cosmological analysis 
\citep{eriksen:2004b} as it contains considerable foreground 
residuals, especially within the Galactic plane, and has a complex pixel 
noise structure.  While these effects were ignored in previous
studies (\eg\ \citealt{bridges:2006b}) since they are sub-dominant on
the large scales of interest when studying Bianchi models, a better
approach is to analyse partial-sky observations of individual \wmap\
frequency bands.

\begin{figure}
\centering
\subfigure[Foreground-reduced W-band map with KQ75 mask applied\label{fig:wmap_data:wband}]{\includegraphics[width=0.45\textwidth]{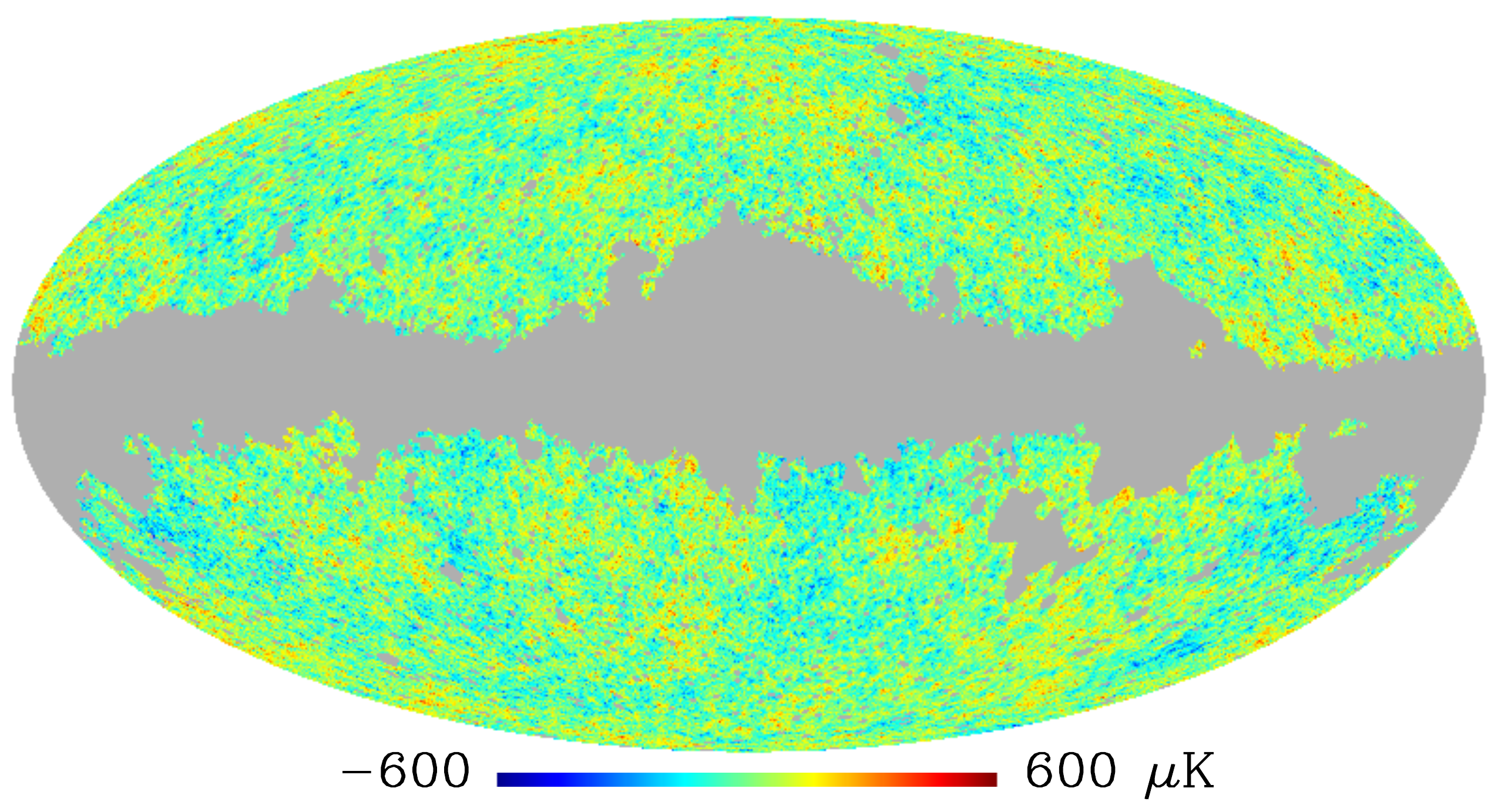}}\quad\quad
\subfigure[ILC map \label{fig:wmap_data:ilc}]{\includegraphics[width=0.45\textwidth]{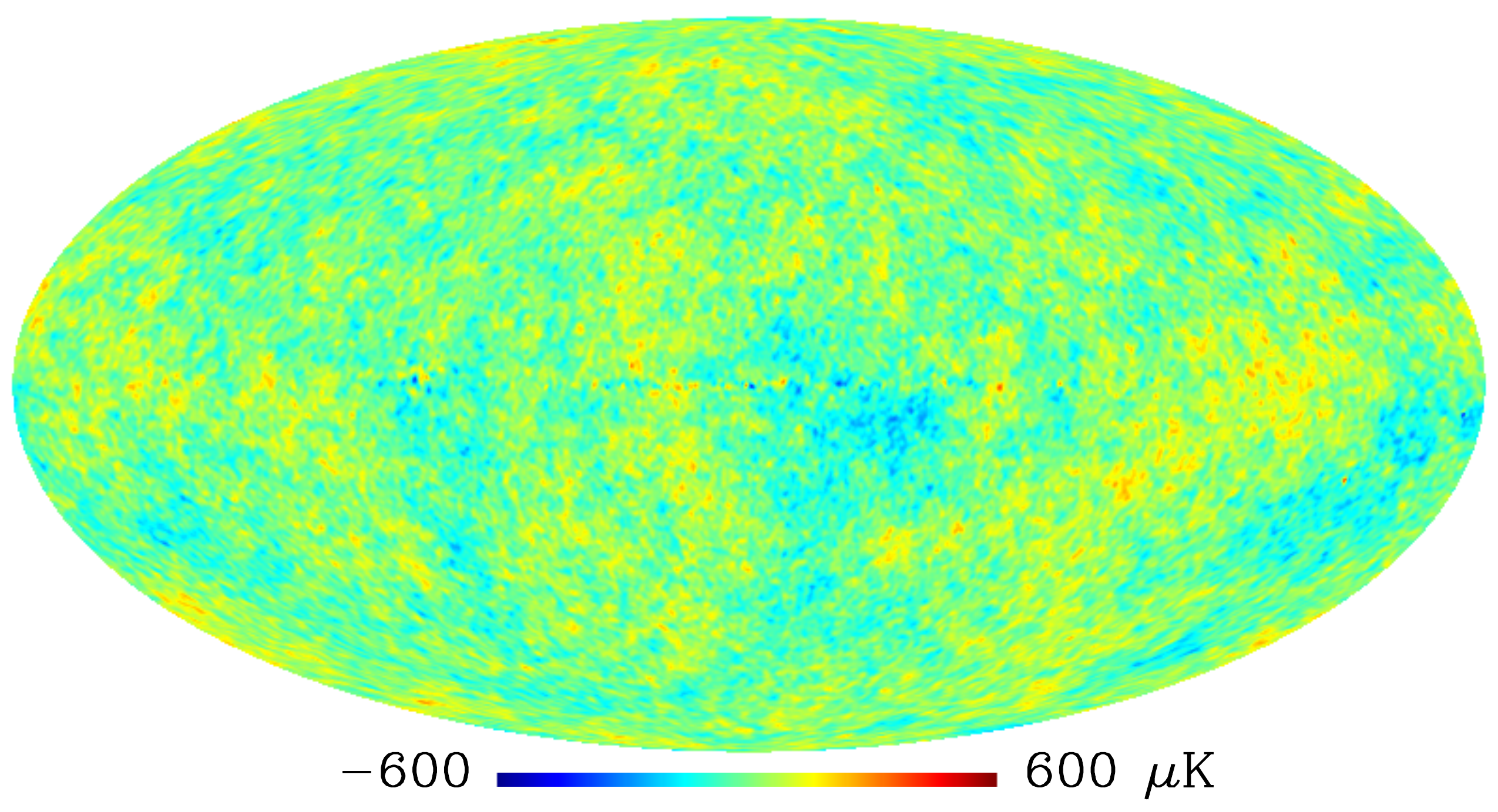}}
\caption{\wmap\ 9-year data analysed.}
\label{fig:wmap_data}
\end{figure}

Our formalism, presented in \sectn{\ref{sec:bayesian_analysis:partial_sky}}, 
allows partial-sky data to be analysed for the first time in a statistically 
rigorous manner, minimising the residual contamination 
in the data and ensuring our conclusions are robust. We analyse the 
foreground-reduced \wmap\ 9-year W-band map \citep{bennett:2013}, 
masked with the conservative 9-year KQ75 sky cut \citep{bennett:2013} 
to excise residual contamination. We select W-band data since this 
band, along with the V-band (which has lower resolution), suffers 
from the least foreground contamination \citep{bennett:2013}. The 
resulting data are shown in \fig{\ref{fig:wmap_data:wband}}. In order 
to draw comparisons with previous studies we also analyse the full-sky 
\wmap\ 9-year ILC map \citep{bennett:2013} shown in 
\fig{\ref{fig:wmap_data:ilc}}, under the proviso that any cosmological 
conclusions drawn from the ILC analysis must be treated with care.

Since \bianchi\ models have a low harmonic band-limit (see \fig{1} of \citealt{mcewen:2006:bianchi}), we use only
low-\el\ \wmap\ data in computing the Bianchi likelihood defined by
\eqn{\ref{eqn:bianchi_likelihood}}; specifically, we compute
contributions to the likelihood of \eqn{\ref{eqn:bianchi_likelihood}}
for $\el\leq\el_{\rm max}^{\rm B}=32$. For the partial-sky setting,
where a non-diagonal covariance matrix must be inverted for each
likelihood evaluation, a low band-limit $\el_{\rm max}^{\rm B}$ is
also convenient to reduce the computational cost of the analysis.  The
\wmap\ data so far described (the ILC map or the KQ75-masked W-band
map in the full- and partial-sky settings respectively) are used for
the Bianchi likelihood evaluation.  As mentioned previously, we
augment the low-\el\ Bianchi likelihood with the standard high-\el\
\wmap\ likelihood \citep{bennett:2013} for $\el>\el_{\rm max}^{\rm B}=32$; this likelihood
function makes use of all \wmap\ temperature observations, as well as
accurate beam models.  The final
log-likelihood is thus the sum of the low-\el\ Bianchi and high-\el\
9-year \wmap\ log-likelihood contributions (since the low- and high-\el\
data are essentially independent, the log-likelihood contributions can
effectively by summed).  As highlighted in 
\sectn{\ref{sec:bayesian_analysis:implementation}}, by
including the \mbox{high-\el} \wmap\ likelihood we are able to constrain
cosmological parameters to greater precision.

We incorporate approximate \wmap\ noise and beam effects in the
low-\el\ Bianchi likelihood.  Since the Bianchi likelihood is
specified in harmonic space (due to the efficiency and accuracy for
which \bianchiviih\ models can be handled in harmonic space), we
approximate the anisotropic \wmap\ noise by isotropic noise (which may
be handled easily in harmonic space due to its diagonal covariance
structure).  We assume isotropic white noise specified by its power
spectrum $N_\el = 0.015\, (\mu {\rm \kelvin})^2$ (computed by $N_\el=
\pixarea \sigma_0^2 / {\rm median}(N_{\rm obs})$, where $N_{\rm obs}$
is the W-band observation-count map and $\sigma_0=6544\,\mu{\rm
  \kelvin}$ is the W-band pixel noise level) and a Gaussian beam with
FWHM of $13.2\arcmin$ in order to approximate W-band observations
accurately. As the individual WMAP bands are smoothed to a common 
resolution of $1^\circ$ when creating the ILC \citep{bennett:2013}, 
we substitute a Gaussian beam of FWHM $1^\circ$ when considering ILC data.
 Since we restrict to low-\el\ data for the Bianchi
likelihood ($\el\leq\el_{\rm max}^{\rm B}=32$), where \wmap\ noise and
beam effects are sub-dominant, these approximations are accurate.  For
the high-\el\ contribution to the likelihood, the official \wmap\
9-year likelihood is used, where noise and beams are modelled very
accurately.

\subsection{Models}
\label{sec:wmap_analysis:models}

We consider three scenarios in this study: one physically motivated, 
and two phenomenological but motivated by previous analyses.

In the first (and only physically motivated) scenario, the Bianchi and
cosmological parameters are coupled; \ie\ the matter and dark-energy 
densities of the Bianchi and standard cosmological models are shared and
thus identical.  Since \bianchiviih\ models are open (with flat models
as the limiting case) we label this model the \emph{open-coupled-Bianchi}
model.  When quoting final conclusions about \bianchiviih\ cosmologies
we will refer to this model since it is a physically well-motivated
and consistent model.

It is also interesting to consider Bianchi \cmb\ contributions 
arising from phenomenological models in which the Bianchi parameters 
are decoupled from the standard cosmological parameters; indeed, 
it is necessary to do so in order to compare to existing results, as this 
is the only approach employed in all previous analyses.
In this non-physical scenario we consider a standard 
\lcdm\ cosmology and a \bianchiviih\ component with
decoupled parameters (\ie\ the total energy density of
the standard cosmology and Bianchi cosmology may differ and are
effectively separate parameters).  As the \lcdm\ cosmologies considered
in previous analyses (\eg\ \citealt{jaffe:2005,bridges:2006b}) were flat, we similarly restrict our attention 
to flat \lcdm\ components, and label this model the
\emph{flat-decoupled-Bianchi} model. Note that in this model we
allow all \lcdm\ parameters to vary.

Finally, we also consider the situation where the majority of the cosmological
parameters are fixed and are not fitted simultaneously with the
\bianchiviih\ parameters.  Following \cite{bridges:2006b}, we do 
allow the amplitude of the primordial power spectrum, $A_s$, to 
vary in this model, so that the amplitude of standard stochastic 
\cmb\ temperature fluctuations is allowed to vary as we fit for 
an embedded Bianchi component. The remaining cosmological 
parameters are fixed at their values as constrained by \wmap\ 
\mbox{9-year} observations, baryon acoustic oscillations and 
supernovae observations \citep{hinshaw:2013}. This is again a
decoupled, non-physical scenario, and we label this model the 
\emph{fixed-decoupled-Bianchi} model, where `fixed' indicates 
that the standard cosmological parameters are essentially fixed.

For each of the three models discussed above, we consider models where
a \bianchiviih\ component is, and is not, included.  Moreover, for the
case where a Bianchi component is included, we consider both
left-handed (LH) and right-handed (RH) Bianchi models, since the
handedness of the coordinate system is also free in \bianchiviih\
models.  Consequently, we recover nine different models that we
analyse (six models that include a Bianchi component and three that do not).

The prior ranges adopted on the \lcdm\ and Bianchi 
parameters in this analysis are shown in \tbl{\ref{tbl:priors}}, and 
are chosen conservatively to reflect our weak prior knowledge of 
the Bianchi models. The priors are uniform in all parameters of 
interest apart from the prior on the power spectrum amplitude, 
$A_s$, which is uniform in $\log A_s$, and the prior on the Euler 
angle \eulb, which is uniform in $\sin$ \eulb. The priors on the 
Bianchi densities, $\Denmat^{\rm B}$ and $\Denlambda^{\rm B}$, 
are applied only in the decoupled models; in the coupled model 
these parameters are set by the sampled \lcdm\ densities
according to $\Denmat^{\rm B} = \Denmatb + \Denmatc$ and 
$\Denlambda^{\rm B} = \Denlambda$.

\begin{table}
  \caption{The prior ranges used for the \lcdm\ and \bianchiviih\
    parameters.}
\label{tbl:priors}
\centering
\begin{tabular}{lc}\toprule
Parameter & Prior Range \\ \midrule
$A_s$ & $[1, 5] \times 10^{-9}$ \\
$n_s$ & $[0.8, 1.1]$ \\
$\tau$ & $[0.082, 0.092]$ \\
$\Denmatb \hubsmall^2$ & $[0.005, 0.05]$ \\
$\Denmatc \hubsmall^2$ & $[0.05, 0.3]$ \\
$\Denlambda$ & $[0.5, 0.9]$ \\
$\Dencurvature$ & $[0.001, 0.2]$ \\ \midrule
\bx & $[0.01, 1]$ \\
\bvortinline & $[0, 1] \times 10^{-10}$ \\
\eula & $[0, 360]^\circ$ \\
\eulb & $[0, 180]^\circ$ \\
\eulc & $[0, 360]^\circ$ \\ \midrule
$\Denmat^{\rm B}$ & $[0, 0.99]$ \\
$\Denlambda^{\rm B}$ & $[0, 0.99]$ \\
\bottomrule
\end{tabular}
\end{table}

\subsection{Results}

We use \wmap\ 9-year data, as specified in
\sectn{\ref{sec:wmap_analysis:data}}, to study the \bianchiviih\
models described in \sectn{\ref{sec:wmap_analysis:models}}, using both
the full- and partial-sky Bayesian analysis techniques presented in
\sectn{\ref{sec:bayesian_analysis}}. We first consider the Bayes
factors computed for different models, before studying parameter
estimates, best-fit maps, and finally vorticity bounds.

\subsubsection{Bayesian evidence}

\begin{table}
  \caption{Log-Bayes factors computed for different models. The Bayes
    factor for each model is computed relative to the corresponding model
    that does not include a Bianchi component, where a positive Bayes factor
    favours the model that does include a Bianchi component.}
\label{tbl:evidences}
\centering
\begin{tabular}{lcc}\toprule
Model & Full-sky ILC & Masked W-band \\ \midrule
Open-coupled-Bianchi (LH) & -0.3 $\pm$ 0.2 & 0.0 $\pm$ 0.2 \\
Open-coupled-Bianchi (RH) & -0.3 $\pm$ 0.2 & 0.1 $\pm$ 0.2 \\
Flat-decoupled-Bianchi (LH) & 1.1 $\pm$ 0.2 & 0.1 $\pm$ 0.2 \\
Flat-decoupled-Bianchi (RH) & 0.1 $\pm$ 0.2 & 0.1 $\pm$ 0.2 \\
Fixed-decoupled-Bianchi (LH) & 1.7 $\pm$ 0.1 & 0.6 $\pm$ 0.1 \\
Fixed-decoupled-Bianchi (RH) & 0.6 $\pm$ 0.1 & 0.3 $\pm$ 0.1 \\
\bottomrule
\end{tabular}
\end{table}

The log-Bayes factors computed for the different models considered are
shown in \tbl{\ref{tbl:evidences}}.  Neither left- nor
right-handed open-coupled-Bianchi models are favoured by \wmap\
9-year data, for either full-sky ILC data or partial-sky W-band data.
Since this is the only physical model studied, we may already conclude
that \wmap\ data do not favour a \bianchiviih\ cosmology over the standard
\lcdm\ cosmology.

Nevertheless, we consider the other phenomenological models discussed
in \sectn{\ref{sec:wmap_analysis:models}} where the Bianchi parameters
are decoupled from the standard cosmological parameters,
since these are the models that have been studied in previous
analyses.  For both the left-handed, flat-decoupled-Bianchi and
fixed-decoupled-Bianchi models, we find evidence in favour of a
Bianchi component (classified as significant on the Jeffreys scale),
when analysing full-sky ILC data. However, when analysing masked
W-band data we find no evidence for these models, suggesting that the
evidence found in the full-sky setting is driven predominantly by
ILC data near the Galactic plane.  No evidence is found for the
corresponding right-hand models.

\subsubsection{Parameter estimates}

\begin{figure*}
\includegraphics[width=0.9\textwidth]{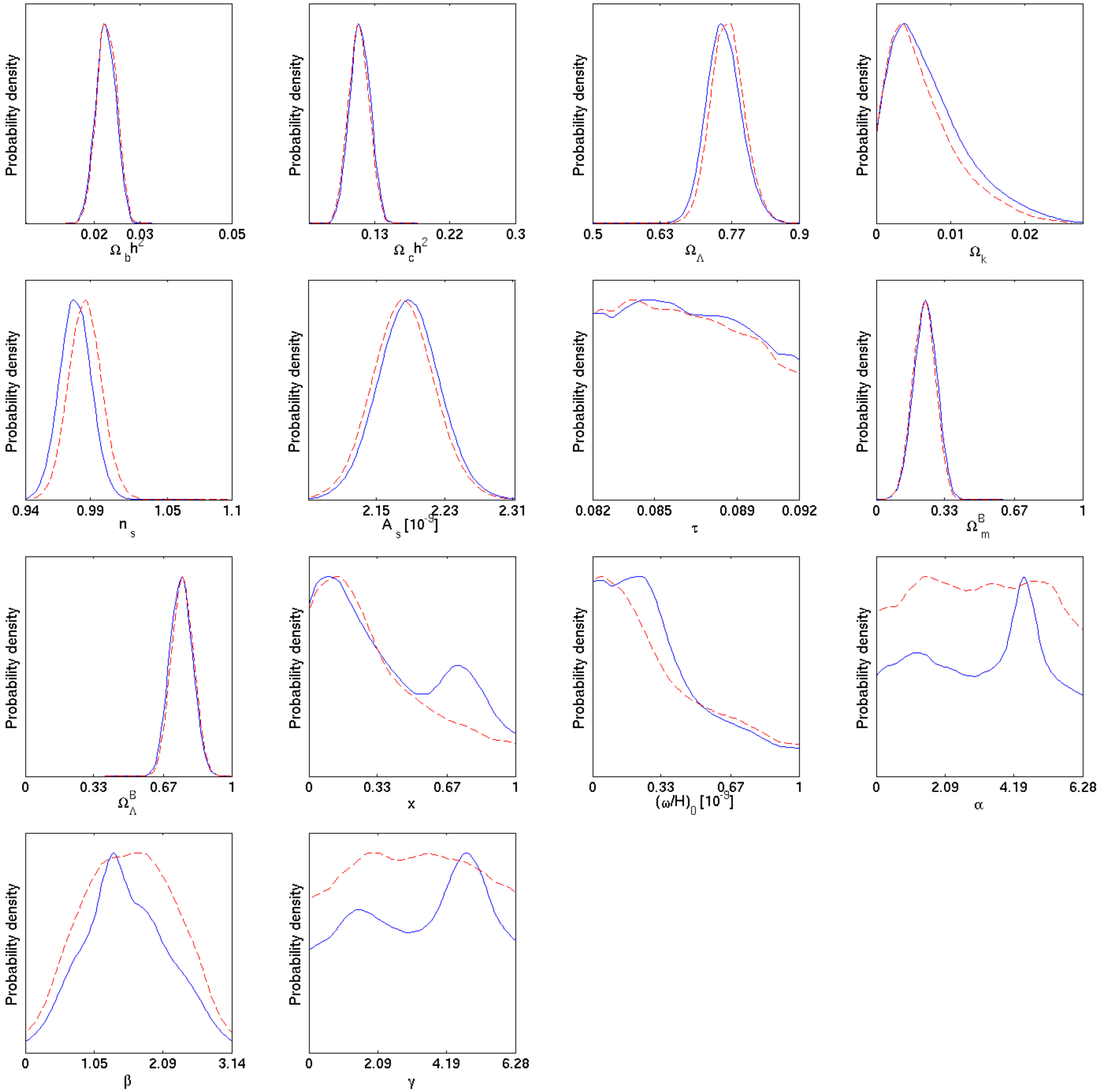}
\caption{Posterior distributions for parameters of the left-handed open-coupled-Bianchi model computed from full-sky ILC data (solid blue curves) and KQ75-masked W-band data (dashed red curves).}
\label{fig:posterior_opencoupledbianchi}
\end{figure*}

\begin{figure*}
\includegraphics[width=0.9\textwidth]{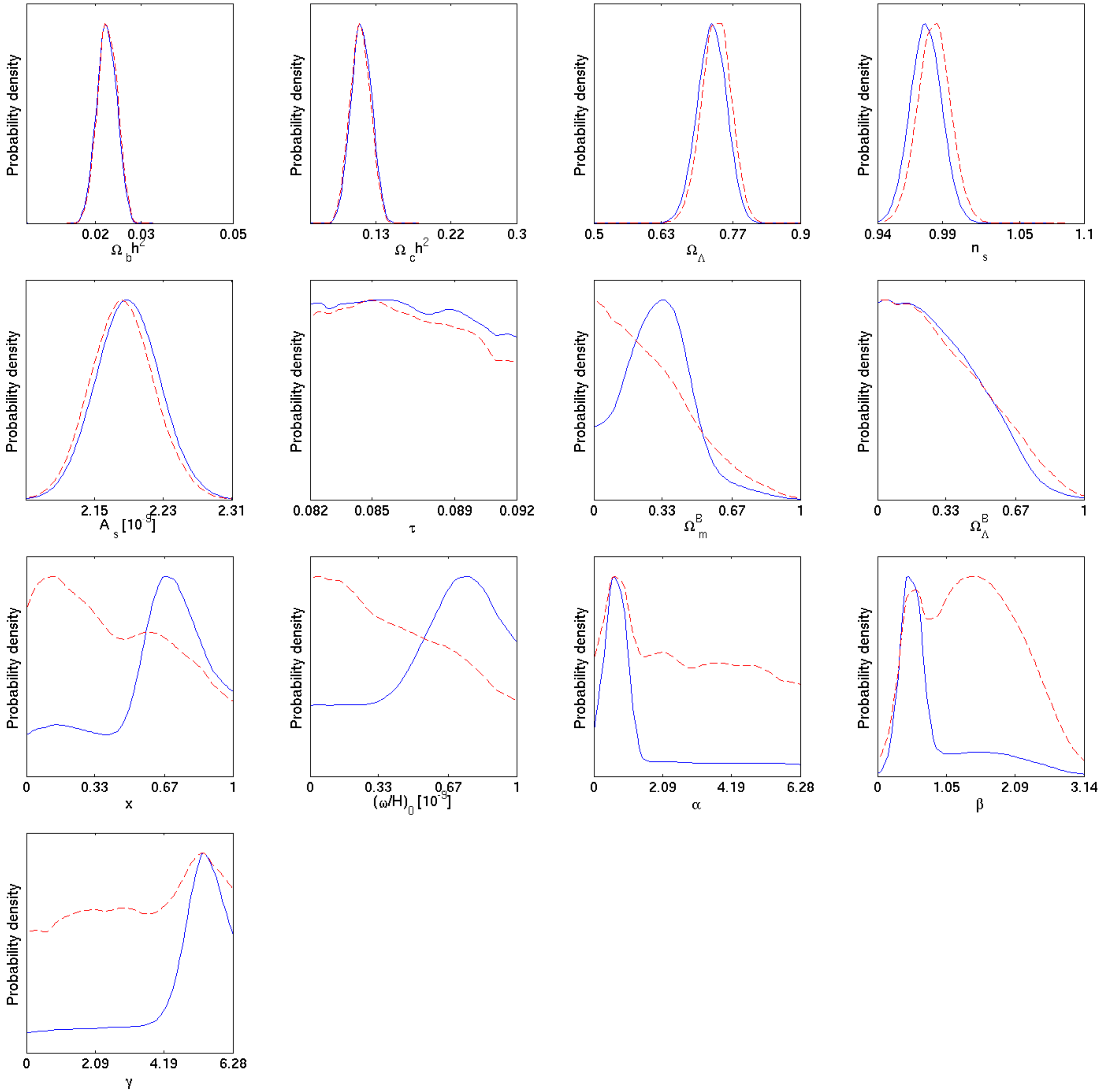}
\caption{Posterior distributions for parameters of the left-handed flat-decoupled-Bianchi model computed from full-sky ILC data (solid blue curves) and KQ75-masked W-band data (dashed red curves).}
\label{fig:posterior_flatdecoupledbianchi}
\end{figure*}

\begin{figure*}
\includegraphics[width=0.9\textwidth]{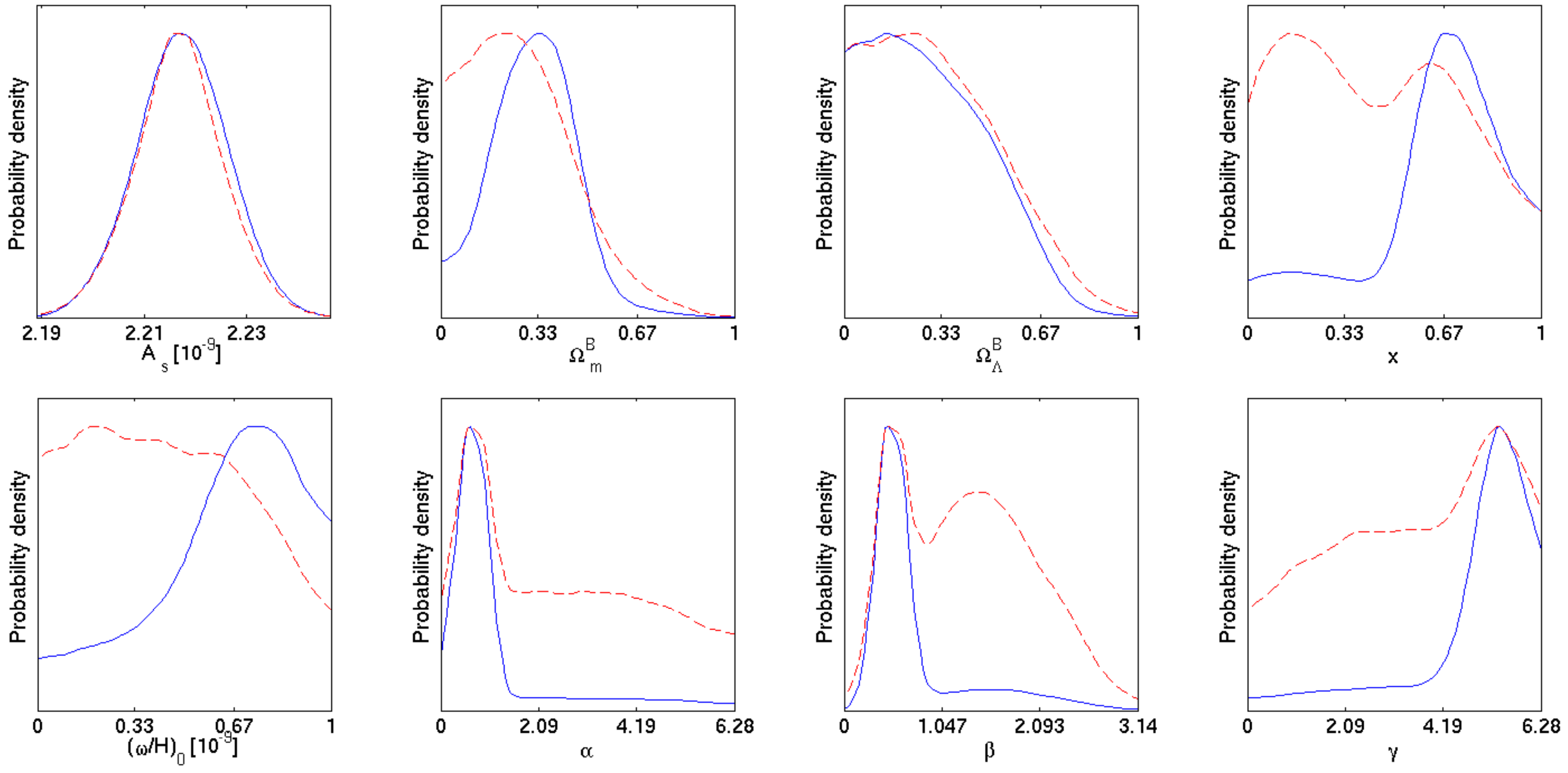}
\caption{Posterior distributions for parameters of the left-handed fixed-decoupled-Bianchi model computed from full-sky ILC data (solid blue curves) and KQ75-masked W-band data (dashed red curves).}
\label{fig:posterior_fixeddecoupledbianchi}
\end{figure*}

Although only left-handed, decoupled-Bianchi models show evidence for
a Bianchi component when analysing full-sky ILC data, for completeness
we show posterior distributions of the parameters of all
left-handed models, using both full-sky ILC data and masked W-band data.
We focus on left-handed models since neither dataset shows evidence
for any right-handed models.

The posterior distributions of the parameters of the
open-coupled-Bianchi model are shown in
\fig{\ref{fig:posterior_opencoupledbianchi}} for full- and
partial-sky data. The standard cosmological parameters are
relatively well constrained thanks to the high-\el\ \wmap\ likelihood
(except for the optical depth to reionisation, $\tau$, as expected
as only \cmb\ temperature data are used). Since the constraints on
the standard cosmological parameters are driven largely by the
high-\el\ \wmap\ likelihood and are similar for all models, we do not
comment on these further when discussing subsequent models.  The
Bianchi parameters, however, are typically poorly constrained for
this model, especially for partial-sky W-band data.  Furthermore,
the Bianchi vorticity \bvortinline, which traces the amplitude of the
Bianchi-induced \cmb\ temperature fluctuations, is peaked near zero.
These posterior plots agree with the conclusion from the Bayesian 
evidence that \wmap\ 9-year data do not favour \bianchiviih\ cosmologies.

The posterior distributions of the parameters of the
flat-decoupled-Bianchi model are shown in
\fig{\ref{fig:posterior_flatdecoupledbianchi}} for full-
and partial-sky data.  For the full-sky analysis, the
Bianchi parameters are relatively well-constrained (except for
$\Denlambda^{\rm B}$ due to the \mbox{$\Denmat^{\rm B}$--$\Denlambda^{\rm
  B}$} degeneracy of \bianchiviih\ models), agreeing with the
previous inference from the Bayesian evidence that a Bianchi component 
is favoured in this setting.  Bianchi parameters are not
well constrained for partial-sky data and, moreover, the Bianchi
vorticity \bvortinline\ is peaked near zero, agreeing with the
conclusion that partial-sky data do not favour a Bianchi component.

The posterior distributions of the parameters of the
fixed-decoupled-Bianchi model are shown in
\fig{\ref{fig:posterior_fixeddecoupledbianchi}} for full-
and partial-sky data.  The conclusions drawn from the
posterior distributions of the parameters of this model are identical
to the flat-decoupled-Bianchi model: essentially, the Bianchi parameters
are well constrained when full-sky ILC data are used, but not when
partial-sky data are used.  Again, these findings agree with conclusions
drawn previously using the Bayesian evidence.

Finally, we summarise the parameter estimates made from the joint
posterior distribution $\prob(\param \given \fitdata, \fitmodelsel)$
for each model in \tbl{\ref{tbl:bianchi_bestfit_parameters_fullsky}}
and \tbl{\ref{tbl:bianchi_bestfit_parameters_partialsky}}, for
partial- and full-sky data respectively. We compute both maximum
\emph{a-posteriori} (MAP) parameter estimates from the peak of the
joint posterior distribution and estimates from the mean of the
marginalised posterior distribution for each parameter. 
One-standard-deviation errors are also quoted for the mean-posterior 
parameter estimates.

\newcommand{\missingparam}{---}
\begin{table*}
  \caption{Parameter estimates recovered for various left-handed
    models from full-sky ILC data. Note that some of these models are not 
      favoured by the Bayesian evidence and some parameters are not well
      constrained.  We nevertheless show all parameter estimates for completeness.}
\label{tbl:bianchi_bestfit_parameters_fullsky}
\centering
\begin{tabular}{cccccccc} \toprule
Parameter     & \multicolumn{2}{c}{Open-coupled-Bianchi} & \multicolumn{2}{c}{Flat-decoupled-Bianchi} & \multicolumn{2}{c}{Fixed-decoupled-Bianchi}\\
              & MAP & Mean & MAP & Mean & MAP & Mean\\ \midrule
$A_s$                  & $2.21\times10^{-9}$ & $(2.19 \pm 0.03)\times10^{-9}$ 
                       & $2.17\times10^{-9}$ & $(2.19 \pm 0.04)\times10^{-9}$ 
                       & $2.219\times10^{-9}$ & $(2.217 \pm 0.009)\times10^{-9}$  \\
$n_s$                  & $0.98$ & $0.98 \pm 0.01$ 
                       & $0.98$ & $0.98 \pm 0.01$ 
                       & \missingparam & \missingparam  \\
$\tau$                 & $0.084$ & $0.087 \pm 0.003$ 
                       & $0.084$ & $0.087 \pm 0.003$ 
                       & \missingparam & \missingparam  \\
$\Denmatb \hubsmall^2$ & $0.023$ & $0.0226 \pm 0.0004$ 
                       & $0.023$ & $0.0225 \pm 0.0004$ 
                       & \missingparam & \missingparam  \\
$\Denmatc \hubsmall^2$ & $0.12$ & $0.112\pm0.004$ 
                       & $0.11$ & $0.112 \pm 0.004$ 
                       & \missingparam & \missingparam  \\
$\Denlambda$           & $0.72$ & $0.75\pm0.03$ 
                       & $0.74$ & $0.73 \pm 0.02$ 
                       & \missingparam & \missingparam  \\
$\Dencurvature$        & $0.002$ & $0.007\pm0.005$ 
                       & \missingparam & \missingparam
                       & \missingparam & \missingparam  \\
$\Denmat^{\rm B}$       & $0.27$ & $0.24\pm0.03$ 
                       & $0.3$ & $0.3 \pm 0.2$ 
                       & $0.4$ & $0.3 \pm 0.1$  \\
$\Denlambda^{\rm B}$    & $0.72$ & $0.75\pm0.03$ 
                       & $0.3$ & $0.3 \pm 0.2$  
                       & $0.2$ & $0.3\pm0.2$  \\
\bx                    & $0.3$ & $0.4 \pm 0.3$
                       & $0.7$&  $0.6 \pm 0.2$  
                       & $0.6$ & $0.6 \pm 0.2$  \\
\bvortinline           & $8 \times 10^{-10}$ & $(3 \pm 2) \times 10^{-10}$ 
                       & $10 \times 10^{-10}$& $(6 \pm 3)  \times 10^{-10}$
                       & $9 \times 10^{-10}$ & $(6 \pm 2) \times 10^{-10}$   \\
\eula                  & $2.5^\circ$ & $181.8^\circ \pm 100.1^\circ$ 
                       & $41.3^\circ$ & $88.8 ^\circ \pm 90.0 ^\circ$  
                       & $41.0^\circ$ & $71.2^\circ \pm 73.7^\circ$ \\
\eulb                  & $122.5^\circ$ & $87.9^\circ \pm 35.7^\circ$ 
                       & $27.2^\circ$ & $50.8^\circ \pm 36.4^\circ$  
                       & $27.6^\circ$ & $43.6^\circ \pm 30.9^\circ$ \\
\eulc                  & $242.5^\circ$ & $189.7^\circ \pm 100.9^\circ$ 
                       & $319.8^\circ$& $259.9^\circ \pm 90.1^\circ$  
                       & $314.1^\circ$ & $277.7^\circ \pm 75.6^\circ$ \\
\bottomrule
\end{tabular}
\end{table*}

\begin{table*}
\caption{Parameter estimates recovered for various left-handed models
  from KQ75-masked W-band data. Note that these models are not
  favoured by the Bayesian evidence and some parameters are not well
  constrained.  We nevertheless show all parameter estimates for completeness.}
\label{tbl:bianchi_bestfit_parameters_partialsky}
\centering
\begin{tabular}{cccccccc} \toprule
Parameter     & \multicolumn{2}{c}{Open-coupled-Bianchi} & \multicolumn{2}{c}{Flat-decoupled-Bianchi} & \multicolumn{2}{c}{Fixed-decoupled-Bianchi}\\
              & MAP & Mean & MAP & Mean & MAP & Mean\\ \midrule
$A_s$                  & $2.17\times10^{-9}$ & $(2.18 \pm 0.04)\times10^{-9}$ 
                       & $2.19\times10^{-9}$ & $(2.18 \pm 0.04)\times10^{-9}$ 
                       & $2.226\times10^{-9}$ & $(2.217 \pm 0.008)\times10^{-9}$  \\
$n_s$                  & $0.98$ & $0.99 \pm 0.01$ 
                       & $0.98$ & $0.98 \pm 0.01$ 
                       & \missingparam & \missingparam  \\
$\tau$                 & $0.084$ & $0.087 \pm 0.003$ 
                       & $0.084$ & $0.087 \pm 0.003$ 
                       & \missingparam & \missingparam  \\
$\Denmatb \hubsmall^2$ & $0.023$ & $0.0228 \pm 0.0005$ 
                       & $0.023$ & $0.0227 \pm 0.0004$ 
                       & \missingparam & \missingparam  \\
$\Denmatc \hubsmall^2$ & $0.11$ & $0.110 \pm 0.004$ 
                       & $0.11$ & $0.111 \pm 0.004$ 
                       & \missingparam & \missingparam  \\
$\Denlambda$           & $0.75$ & $0.76 \pm 0.03$ 
                       & $0.73$ & $0.74 \pm 0.02$ 
                       & \missingparam & \missingparam  \\
$\Dencurvature$        & $0.003$ & $0.007\pm0.005$ 
                       & \missingparam & \missingparam
                       & \missingparam & \missingparam  \\
$\Denmat^{\rm B}$       & $0.25$ & $0.23 \pm 0.03$ 
                       & $0.3$ & $0.3 \pm 0.2$ 
                       & $0.1$ & $0.3 \pm 0.2$  \\
$\Denlambda^{\rm B}$    & $0.75$ & $0.76 \pm 0.03$ 
                       & $0.4$ & $0.3 \pm 0.2$  
                       & $0.4$ & $0.3 \pm 0.2$  \\
\bx                    & $0.5$ & $0.3 \pm 0.3$ 
                       & $0.7$ & $0.4 \pm 0.3$  
                       & $0.2$ & $0.5\pm 0.3$  \\
\bvortinline           & $6 \times 10^{-10}$ & $(3 \pm 3) \times 10^{-10}$ 
                       & $8 \times 10^{-10}$& $(4 \pm 3)  \times 10^{-10}$
                       & $4 \times 10^{-10}$ & $(4 \pm 3) \times 10^{-10}$ \\
\eula                  & $55.6^\circ$ & $179.0^\circ \pm 101.0^\circ$ 
                       & $37.0^\circ$ & $162.8^\circ \pm 104.3^\circ$  
                       & $244.0^\circ$ & $150.2^\circ \pm 102.5^\circ$ \\
\eulb                  & $59.3^\circ$ & $90.6^\circ \pm 38.3^\circ$ 
                       & $26.4^\circ$ & $82.1^\circ \pm 39.6^\circ$  
                       & $106.9^\circ$ & $75.3^\circ \pm 39.0^\circ$ \\
\eulc                  & $102.6^\circ$ & $180.5^\circ \pm 100.3^\circ$ 
                       & $321.0^\circ$ & $193.2^\circ \pm 103.8^\circ$  
                       & $170.3^\circ$ & $203.8^\circ \pm  100.3^\circ$ \\
\bottomrule
\end{tabular}
\end{table*}

\subsubsection{Best-fit maps}

\begin{figure*}
\centering
\subfigure[Flat-decoupled-Bianchi model \label{fig:bestfit_maps:flatdecoupledbianchi}]{\includegraphics[width=0.45\textwidth]{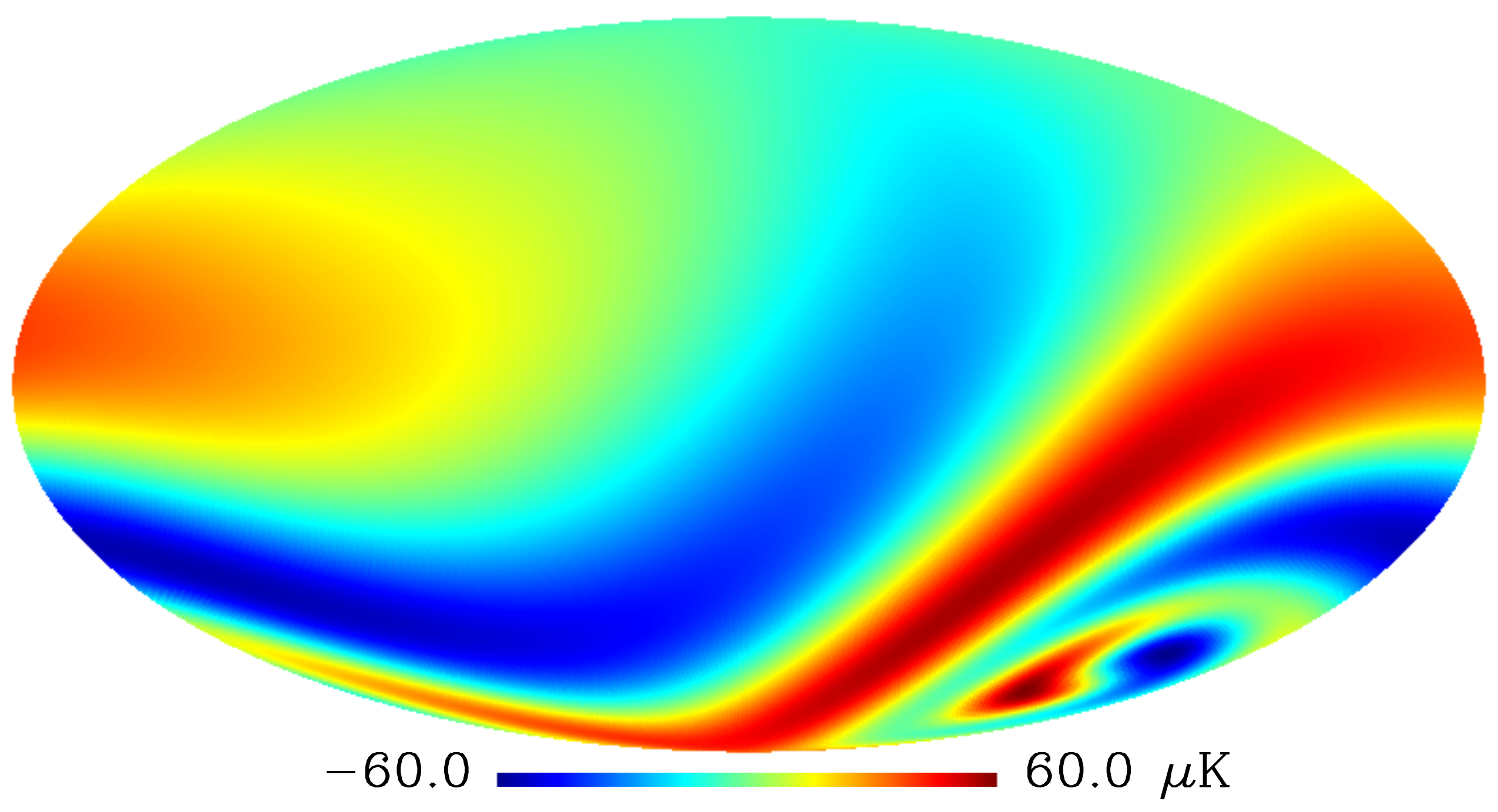}}\quad\quad
\subfigure[Fixed-decoupled-Bianchi model \label{fig:bestfit_maps:fixeddecoupledbianchi}]{\includegraphics[width=0.45\textwidth]{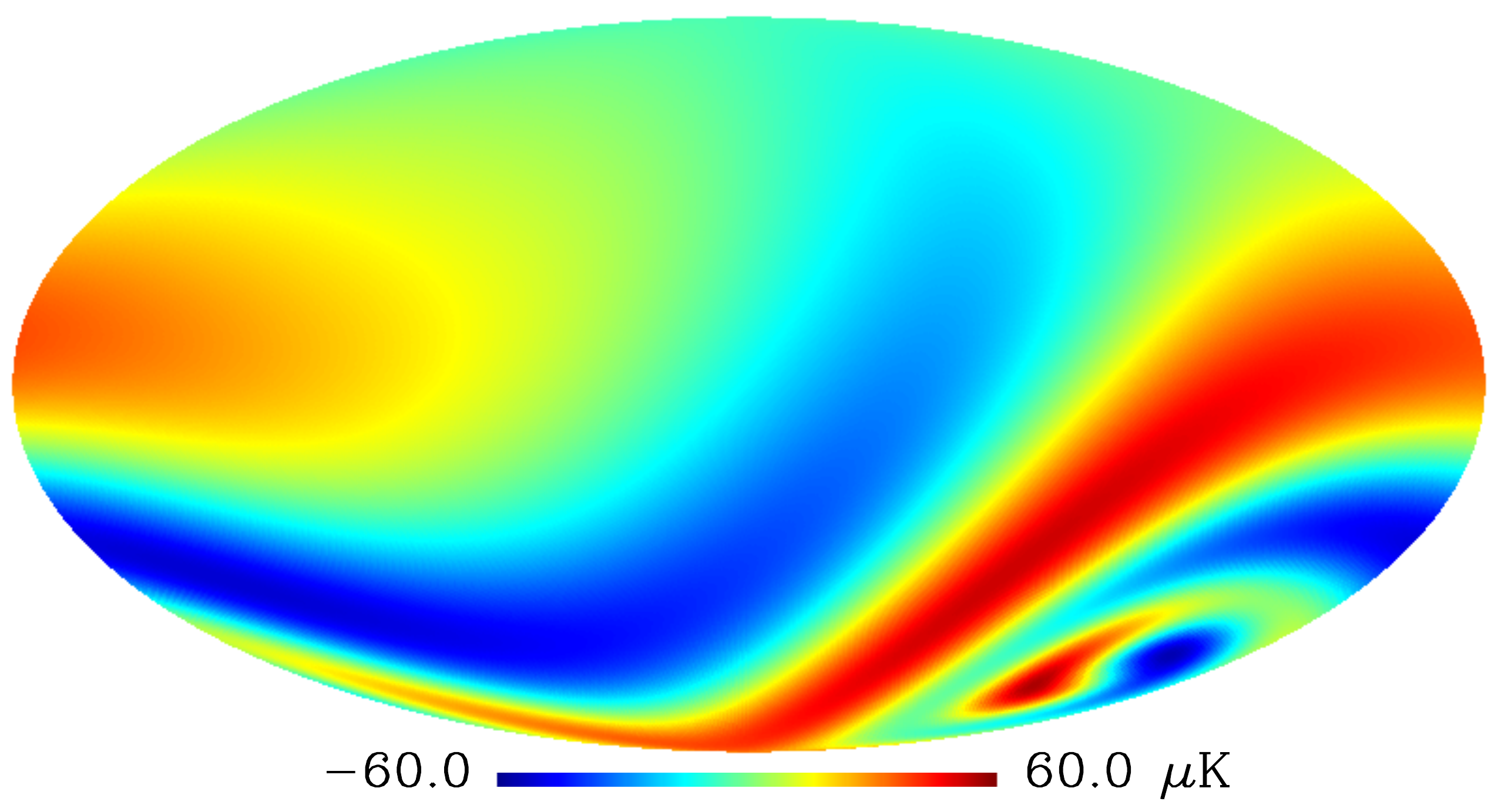}}
\caption{Best-fit non-physical \bianchiviih\ templates found in full-sky ILC data.}
\label{fig:bestfit_maps}
\end{figure*}

For the two models where the Bayes factor provides significant
evidence in support of a Bianchi component (see
\tbl{\ref{tbl:evidences}}), namely the non-physical
flat-decoupled-Bianchi and fixed-decoupled-Bianchi models, we plot the
best-fit embedded Bianchi maps in
\fig{\ref{fig:bestfit_maps:flatdecoupledbianchi}} and
\fig{\ref{fig:bestfit_maps:fixeddecoupledbianchi}}, respectively.
Best-fit maps are computed from the MAP parameter estimates contained
in \tbl{\ref{tbl:bianchi_bestfit_parameters_fullsky}}.
These best-fit maps are very similar to those found in previous
releases of \wmap\ data \citep{jaffe:2005,bridges:2006b}.

\subsubsection{Vorticity bounds}

We conclude our analysis of \wmap\ 9-year data by reiterating that
only the open-coupled-Bianchi model is physical. The other models
considered are non-physical since the Bianchi parameters are decoupled
from the standard cosmological parameters: the Bianchi component can
therefore require parameters, for example, a total energy density, that are
incompatible with the stochastic CMB component.  For the
physical open-coupled-Bianchi model no evidence for a Bianchi component is found in either full-sky ILC or
partial-sky W-band data.  In this physical model we can therefore place
constraints on the vorticity of \bianchiviih\ cosmologies.  For
full-sky ILC data we recover the constraint $\bvortinline < 8.1 \times
10^{-10}$ at 95\% confidence, for both left- and right-handed models.
For partial-sky W-band data we recover the constraint $\bvortinline <
8.1 \times 10^{-10}$ for the left-handed model and the constraint
$\bvortinline < 8.6 \times 10^{-10}$ for the right-handed model, both
at 95\% confidence.  These constraints on the global anisotropy of our
Universe are placed in the context of a well-motivated, physical
model, namely \bianchiviih\ cosmologies, using robust Bayesian
statistical methods.

\section{Conclusions}
\label{sec:conclusions}

We have performed a definitive analysis of \bianchiviih\ cosmologies
with \wmap\ 9-year temperature data.  We have studied full-sky ILC
data and masked W-band data, using Bayesian analysis techniques
developed for the full- and partial-sky settings. Three different 
models were studied, namely the physically motivated
open-coupled-Bianchi model, where the Bianchi and standard cosmological
parameters are coupled, and the flat-decoupled-Bianchi and
fixed-decoupled-Bianchi models, which are non-physical since the Bianchi
and standard cosmological parameters are decoupled.  Although we have
focused here on Bianchi models, our analysis techniques may be easily
extended to study other anisotropic cosmologies such as non-trivial
topologies; such models will be the focus of future research.  

For the non-physical decoupled models, we find Bayesian evidence
favouring the inclusion of a left-handed Bianchi component when
analysing full-sky ILC data.  The resulting best-fit Bianchi maps
found in \wmap\ 9-year data are similar to those found in previous
releases of \wmap\ data \citep{jaffe:2005,bridges:2006b}.  However,
when studying these models with masked W-band data we find no evidence
for a Bianchi component, suggesting that data near the Galactic
plane provide a large contribution to the evidence found in the
full-sky setting.

For the physical \bianchiviih\ model, we find no evidence for the
inclusion of a Bianchi component in either full- or partial-sky \wmap\
data.  Since this is a well-motivated, physical model we can state
definitively that \wmap\ data do not favour \bianchiviih\ cosmologies
over \lcdm.  However, neither is it possible to conclusively discount
\bianchiviih\ cosmologies in favour of \lcdm\ cosmologies. We
constrain the vorticity of \bianchiviih\ cosmologies, and hence the
large-scale anisotropy of the universe, at $\bvortinline < 8.6 \times
10^{-10}$ with 95\% confidence.

\bianchiviih\ cosmologies induce a global anisotropic \cmb\ temperature
contribution that is inherently low-resolution, so the higher
resolution of forthcoming \emph{Planck} \citep{planck:bluebook} data
as compared to \wmap\ is not in itself relevant to enhanced constraints
on these models.  However, \emph{Planck}'s ability to remove large-scale
astrophysical foregrounds will result in \cmb\ observations of
unprecedented precision over almost the entire sky. An analysis of
\bianchiviih\ cosmologies using \emph{Planck} temperature data will thus be
very informative.  Moreover, polarised \emph{Planck} observations will also
have the potential to further constrain \bianchiviih\ cosmologies.

\section*{Acknowledgements}

We thank Anthony Challinor for useful discussions about masking in
harmonic space.  JDM is supported in part by a Newton International Fellowship
from the Royal Society and the British Academy. TJ was supported by
STFC during the course of this work. SMF is supported by STFC and a
grant from the Foundational Questions Institute (FQXi) Fund, a
donor-advised fund of the Silicon Valley Community Foundation on the
basis of proposal FQXi-RFP3-1015 to the Foundational Questions
Institute.  HVP is supported by STFC, the Leverhulme Trust, and the
European Research Council under the European Community's Seventh
Framework Programme (FP7/2007- 2013) / ERC grant agreement no
306478-CosmicDawn.  We acknowledge use of the following public
software packages: {\tt HEALPix} \citep{gorski:2005}; {\tt CAMB}
\citep{lewis:2000}; {\tt MultiNest}
\citep{feroz:multinest1,feroz:multinest2}; {\tt BIANCHI2}
\citep{mcewen:2006:bianchi}; {\tt S2} \citep{mcewen:2006:fcswt}.  We
acknowledge use of the Legacy Archive for Microwave Background Data
Analysis (LAMBDA).  Support for LAMBDA is provided by the NASA Office
of Space Science.  The authors acknowledge the use of the UCL Legion
High Performance Computing Facility (Legion@UCL), and associated
support services, in the completion of this work.

\bibliographystyle{mymnras_eprint}
\bibliography{bib}

\label{lastpage}
\end{document}